\documentclass[aps, arxiv, floatfix, twocolumn, superscriptaddress]{revtex4-2}

\usepackage[T1]{fontenc}
\usepackage{graphicx}
\usepackage{bm}
\usepackage{amsmath}
\usepackage{amssymb}
\usepackage{mathtools}
\usepackage[version=4]{mhchem}
\usepackage{float}

\usepackage[colorlinks, allcolors=blue]{hyperref}

\usepackage{microtype}

\begin{document}

\title{Scalable Training of Neural Network Potentials for Complex Interfaces \\ Through Data Augmentation}

\author{In Won Yeu}
\affiliation{Department of Chemical Engineering, Columbia University, New York, NY, USA.}
\affiliation{Columbia Center for Computational Electrochemistry, Columbia University, New York, NY, USA.}
\author{Annika Stuke}
\affiliation{Department of Chemical Engineering, Columbia University, New York, NY, USA.}
\affiliation{Columbia Center for Computational Electrochemistry, Columbia University, New York, NY, USA.}
\author{Jon López-Zorrilla}
\affiliation{Physics Department, University of the Basque Country (UPV/EHU), Leioa, Basque Country, Leioa, Spain.}
\author{James M. Stevenson}
\affiliation{Schrödinger, Inc., New York, NY, USA.}
\author{David R. Reichman}
\affiliation{Columbia Center for Computational Electrochemistry, Columbia University, New York, NY, USA.}
\affiliation{Department of Chemistry, Columbia University, New York, NY, USA.}
\author{Richard A. Friesner}
\affiliation{Columbia Center for Computational Electrochemistry, Columbia University, New York, NY, USA.}
\affiliation{Department of Chemistry, Columbia University, New York, NY, USA.}
\author{Alexander Urban}
\email{a.urban@columbia.edu}
\affiliation{Department of Chemical Engineering, Columbia University, New York, NY, USA.}
\affiliation{Columbia Center for Computational Electrochemistry, Columbia University, New York, NY, USA.}
\affiliation{Columbia Electrochemical Energy Center, Columbia University, New York, NY, USA.}
\author{Nongnuch Artrith}
\email{n.artrith@uu.nl}
\affiliation{Columbia Center for Computational Electrochemistry, Columbia University, New York, NY, USA.}
\affiliation{Debye Institute for Nanomaterials Science, Utrecht University, 3584 CS Utrecht, The Netherlands.}

\date{\today}

\begin{abstract}
\noindent
Artificial neural network (ANN) potentials enable highly accurate atomistic simulations of complex materials at unprecedented scales.
Despite their promise, training ANN potentials to represent intricate potential energy surfaces (PES) with transferability to diverse chemical environments remains computationally intensive, especially when atomic force data are incorporated to improve PES gradients.
Here, we present an efficient ANN potential training methodology that uses Gaussian process regression (GPR) to incorporate atomic forces into ANN training, leading to accurate PES models with fewer additional first-principles calculations and a reduced computational effort for training.
Our GPR-ANN approach generates synthetic energy data from force information in the reference dataset, thus augmenting the training datasets and bypassing direct force training.
Benchmark tests on hybrid density-functional theory data for ethylene carbonate (EC) molecules and Li metal-EC interfaces, relevant for lithium metal battery applications, demonstrate that GPR-ANN potentials achieve accuracies comparable to fully force-trained ANNs with a significantly reduced computational overhead.
Detailed comparisons show that the method improves both data efficiency and scalability for complex interfaces and heterogeneous environments.
This work establishes the GPR-ANN method as a powerful and scalable framework for constructing high-fidelity machine learning interatomic potentials, offering the computational and memory efficiency critical for the large-scale simulations needed for the simulation of materials interfaces.
\end{abstract}

\maketitle

\section{Introduction}
\label{sec:introduction}

Interactions at materials interfaces are essential to technologically relevant phenomena, such as crystal growth~\cite{yeu_ab_2020, cheula_local_2021}, catalytic activity~\cite{artrith_understanding_2014}, and interphase formation~\cite{winter2009solid, heiskanen_generation_2019}.
A concrete example is lithium metal batteries, which are a promising alternative to conventional Li-ion batteries due to their potential for higher energy density and lower production cost~\cite{xu_lithium_2014, eroglu_fraction_2014, pang_electrolyteelectrode_2021, famprikis_fundamentals_2019, banerjee_interfaces_2020, xiao_understanding_2019}.
However, their commercialization has been hindered by a lack of understanding regarding the reaction between lithium metal and liquid electrolytes~\cite{li_revealing_2017, yuan_ultrafast_2023}.
An atomistic understanding of interface structures and reaction dynamics would provide an opportunity to control interfaces in devices such as Li-metal batteries.
Unfortunately, experimental characterization of interfaces \emph{in operando} remains challenging, and current simulation approaches either face prohibitive computational costs or lack sufficient accuracy.

Interactions at interfaces typically involve different types of bonding, e.g., metallic bonding within lithium metal and covalent and ionic bonding in the electrolyte, which is not well captured by conventional interatomic potentials.
Additionally, interface simulations typically require structure models with several hundred to thousands of atoms, i.e., system sizes that are challenging for accurate first-principles electronic structure methods.
Semilocal density-functional theory (DFT), which is comparatively efficient and the most widely used electronic structure method for materials simulations, exhibits significant errors if the two materials in contact exhibit different types of bonding, requiring computationally significantly more demanding hybrid functionals to reliably describe the electronic structure and potential energy in interface regions~\cite{debnath_accurate_2023}.

Machine-learning (ML) potentials trained on first-principles and quantum chemistry methods have emerged as a new family of reactive interatomic force fields~\cite{behler2017first, behler_perspective_2016, himanen_data-driven_2019, deringer_machine_2019, schmidt_recent_2019, handley_potential_2010}.
Early methods, including Gaussian Process regression~\cite{bartok2010gaussian} and feed-forward neural networks~\cite{behler2007generalized} with atomic descriptors~\cite{behler_atom-centered_2011}, laid the groundwork for ML potentials, while recent innovations~\cite{shapeev_moment_2016, schutt_schnet_nodate, batzner_e3-equivariant_2022, chen_graph_2019, chen_universal_2022, deng_chgnet_2023, drautz_atomic_2019, batatia_mace_nodate} achieve gradual improvements in terms of computational efficiency and accuracy on public benchmarks, bringing ML potentials closer to replacing DFT.
Especially, ML potentials based on artificial neural networks (ANNs) with atomic descriptors have been applied to a wide range of materials and phenomena, including metals \cite{khaliullin_nucleation_2011, eshet_ab_2010, artrith_high-dimensional_2012, boes_neural_2016, sun_metastable_2018, artrith_neural_2013}, oxides \cite{artrith_implementation_2016, artrith_high-dimensional_2011,artrith_neural_2013, elias_elucidating_2016}, alloys \cite{kobayashi_neural_2017, sosso_neural_2012}, molecular systems \cite{morawietz_how_2016, ssmith_ani-1_2017, cooper_potential_2018, morawietz_hiding_2019} and amorphous phases \cite{li_study_2017,artrith_constructing_2018,deringer_realistic_2018, lacivita_structural_2018} due to their computational efficiency and easy accessibility.
Carefully trained ML potentials can represent the PES of materials with thousands to millions of atoms with an accuracy close to that of \emph{ab initio} methods at a significantly reduced computational cost and scaling.
Accordingly, extensive research has been devoted to constructing ANN potentials for interfaces, for instance, between copper clusters and zinc oxide~\cite{artrith_neural_2013}, water and copper~\cite{natarajan_neural_2016}, water and zinc oxide~\cite{quaranta_maximally_2018}, and heterogeneous catalysts~\cite{omranpour_machine_2024, vandermause_active_2022}.
However, constructing reliable interatomic potentials that are able to represent surfaces and interfaces is especially challenging because abrupt changes in atomic environments and different bonding types are involved.
As a consequence, a huge amount of reference data points can be needed to capture the drastically changing potential energy surface (PES) with sufficient precision~\cite{artrith_machine_2019}.
These challenges have limited the construction of ANNs for interfaces, and atomistic understanding of interfacial reactions remains limited despite their crucial impacts in various technology areas~\cite{zheng_reversible_2019, etxebarria_work_2020, li_understanding_2022}.

Effective learning strategies are desirable to avoid any unnecessary first-principles calculations with expensive hybrid functionals or higher-level theory.
In this regard, including atomic force information in the ANN potential training was found to greatly reduce data requirements, improve PES accuracy, and increase transferability~\cite{witkoskie_neural_2005, marques_neural_2019, pukrittayakamee_simultaneous_2009, gastegger_high-dimensional_2015, behler_atom-centered_2011, artrith_high-dimensional_2011, zhang_deep_2018}.
Additionally, active learning can be employed, where training data for ML potential construction is generated incrementally based on the current state of the potential, enabling adding new data to training sets in a systematic and non-redundant fashion~\cite{jinnouchi_--fly_2020,  yang_using_2022, smith_less_2018, ang_active_2021, zeng_complex_2020, yang_using_2022, ayoung_transferable_2021}.
A typical active learning strategy is to perform additional first-principles calculations for atomic structures for which the ML potential reports an uncertainty that exceeds a user-defined threshold.
Such an approach avoids redundant first-principles calculations and increases the transferability of ML potentials by adapting the model to a new structure domain.

However, training ANNs not only on function values (energies) but also on derivatives (atomic forces, stress tensors, etc.) comes at a significant computational and memory overhead because such direct force training needs to evaluate and store the second (or higher order) derivative of the ANN potential, which scales quadratic with the number of atoms within the cutoff range~\cite{cooper_efficient_2020}.
This unfavorable scaling can be prohibitively expensive for complex, dense systems, or at least calls for expensive specialized hardware.
Furthermore, conventional ANNs do not directly provide an uncertainty estimate that could be used for active learning, so either the predictions from multiple independently-trained ANN potentials need to be combined (\emph{query by committee})~\cite{artrith_high-dimensional_2012, schran_committee_2020} or the ANN architecture needs to be modified, for example, by introducing dropout layers~\cite{gal_dropout_nodate}.

\begin{figure*}[t]
  \centering
  \includegraphics[width=0.9\textwidth]{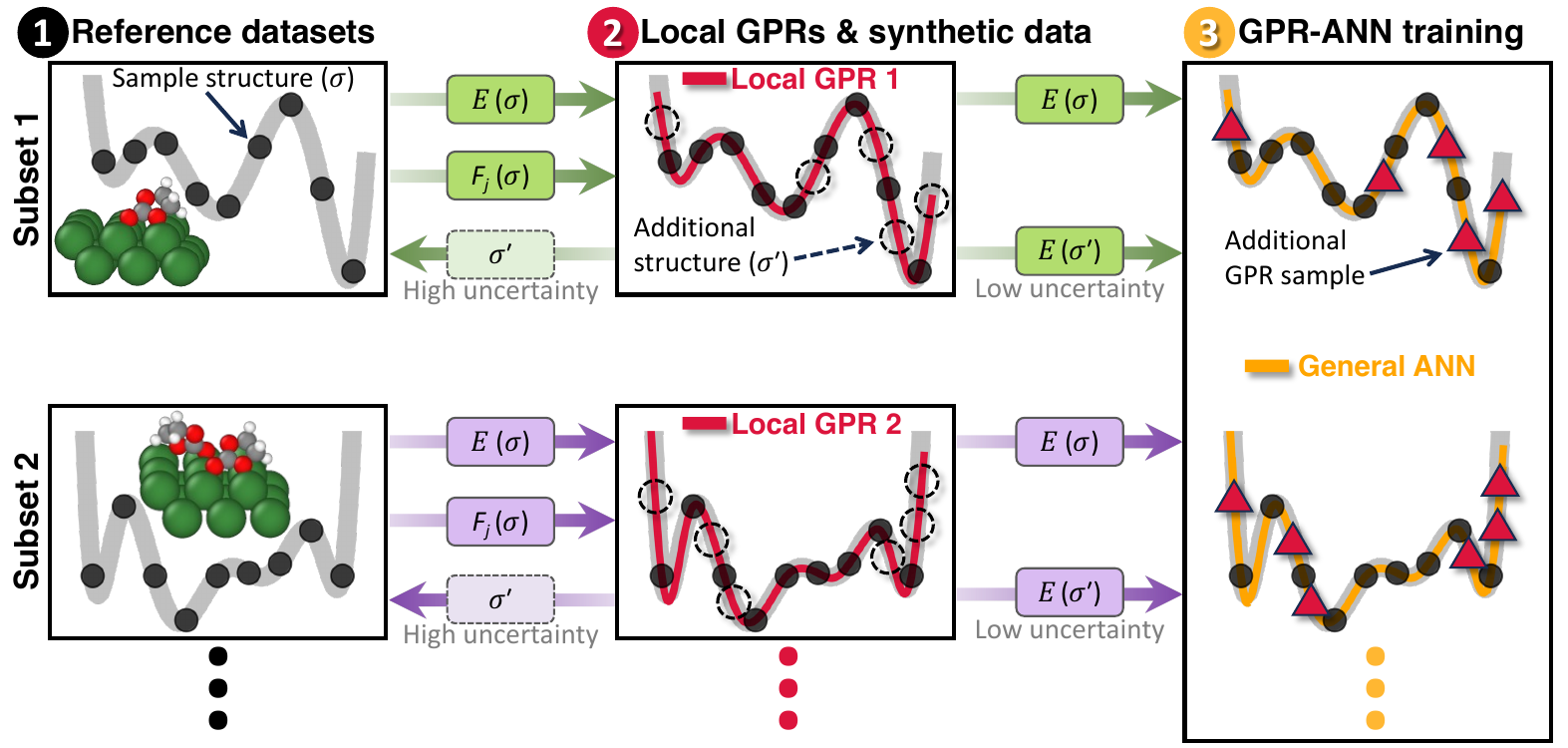}
  \caption{\label{fig:workflow}
    \textbf{Indirect force training with the GPR-ANN approach. } (\textbf{Step~1}) The reference data (black circles) consists of atomic structures ($\sigma$), their energies ($E(\sigma)$) and corresponding atomic forces ($F_{j}(\sigma)$) from electronic structure calculations for structures sampling target potential energy surfaces (PES, thick gray lines).  Each subset contains related structures with the same number of atoms.  (\textbf{Step~2}) For each subset, Gaussian process regression (GPR) models can efficiently interpolate the potential energy surface based on the energies and atomic forces (red lines).  The GPR models can then be used to generate synthetic data by labeling additional related structures (empty circles) with energies.  Structures for which the GPR model reports a high uncertainty are evaluated with the reference electronic structure method.  (\textbf{Step~3}) Finally, the original structures and their energies can be combined with the additional structures and their GPR energies (red triangles) into a unified overall data set that can be used for efficient energy-only training of general ANN potentials (yellow lines).}
\end{figure*}

In this article, we introduce a new data-augmentation approach where ANN training is seamlessly integrated with Gaussian process regression (GPR), a non-parametric regression model, to overcome these downsides of ANN training.
The \emph{GPR-ANN} approach indirectly learns the information from the PES gradients (i.e., the interatomic forces) by translating the gradients to additional energy data via local interpolation and extrapolation using separate GPR models simultaneously fitting to data points and their derivatives of subsystems of overall heterogeneous reference data (\textbf{Figure~\ref{fig:workflow}}).
The general idea follows the same spirit as the first-order Taylor-expansion extrapolation method that some of the present authors proposed previously~\cite{cooper_efficient_2020} and simple extrapolation based on the zeroth order~\cite{gibson_data-augmentation_2022}.
However, the non-linear, Bayesian nature of GPR models such as the Gaussian approximation potential by Bartók et al.~\cite{bartok2010gaussian, bartok_machine_nodate, bartok_machine_2018, bartok2015g}, which are inherently less prone to overfitting in new domains than ANNs and perform exceptionally well with limited small data among various ML potential methods~\cite{kamath_neural_2018, zuo_performance_2020}, leads to greatly improved performance, as we will show in the following.
We show how the GPR-ANN approach enables scalable force training without relying on direct force training by combining the \emph{best of both worlds}: leverages the GPR’s capabilities of superior interpolation and extrapolation with small data sets and the uncertainty estimation at no additional computational cost, which can be used for Bayesian active learning, as surrogate models to augment filtered synthetic data, while mitigating GPR’s prohibitive computational cost for large data sets with efficient ANN training.

In the following \emph{Results} section, we first detail the working principle of the GPR-ANN method and then demonstrate its improved performance in comparison with conventional ANN potentials by applying the method to three benchmark cases with increasing complexity:
(i)~a Lennard-Jones (LJ) potential of the \ce{H-H} bond in the \ce{H2} molecule, and (ii)~a hybrid-functional DFT PES of two ethylene carbonate (EC) molecules, and (iii)~an EC molecule on the surface of lithium (Li) metal.

\section{Results}
\label{sec:results}

\subsection{Energy training}

A popular ANN potential architecture is the high-dimensional PES proposed by Behler and Parrinello~\cite{behler2007generalized}, which describes the total energy, $E(\sigma)$, of a structure, $\sigma=\{(\vec{R}_1, t_1), (\vec{R}_2, t_2), ..., (\vec{R}_N, t_N)\}$ where $\vec{R}_i$ are the coordinates of atom $i$ and $t_i$ is its chemical species, as a sum of atomic energy contributions
\begin{align}\label{eq:ANN-energy}
    E(\sigma)
    \approx E^{\textup{ANN}}(\sigma ; \{w\})
    = \sum_{i\in\sigma}\textup{ANN}_{t_i}(\sigma_i^{R_c} ; \{w_{t_i}\})
\end{align}
where $\sigma_i^{R_c}$ in Equation~\ref{eq:ANN-energy} is a descriptors (i.e., feature vector) representing the atomic environment of atom $i$ within a cutoff radius $R_c$ that serves as input to a multilayer perceptron feedforward neural network, $\textup{ANN}_{t_i}$, specific to the chemical species of atom $i$.
Each neural network $\textup{ANN}_{t_i}$ is defined by its weight parameters $\{w_{t_i}\}$, and we denote the set of all weight parameters for all chemical species $\{w\}$.

A basic requirement of $\sigma_i^{R_c}$ is to obey the invariances of the total energy with respect to translation/rotation of the entire structure and permutation of equivalent atoms, and in this work, we used a Chebyshev descriptor method~\cite{artrith_efficient_2017} that allows for an efficient representation of multi-element compounds.
Details of the ANN architecture and the parameters for the Chebyshev descriptor are given in the \emph{Methods} section.

Given reference data sets of structures {$\sigma$} and energies $E^{\textup{ref}}(\sigma)$, energy-only training minimizes the energy loss function
\begin{align}\label{eq:energy-loss}
  	\mathcal{L}^{\textup{energy}}
   = \sum_{\sigma}\frac{1}{2}\bigl\{E^{\textup{ANN}}(\sigma ; \{w\}) - E^{\textup{ref}}(\sigma) \bigl\}^2
\end{align}
by optimizing the weight parameters $\{w\}$
\begin{align}\label{eq:energy-loss-optimization}
    \{w^{\textup{opt}}\} = \arg\min_{\{w\}}\{\mathcal{L}^{\textup{energy}}\}.
\end{align}
We refer to the process of minimizing the loss function $\mathcal{L}^{\textup{energy}}$ as \emph{energy training}.

The minimization of $\mathcal{L}^{\textup{energy}}$ with respect to $\{w\}$ requires the derivative
\begin{align}\label{eq:energy-loss-derivative}
    \frac{\partial \mathcal{L}^{\textup{energy}}}{\partial w}
    &= \sum_{\sigma} \Delta E(\sigma)\frac{\partial E^{\textup{ANN}}(\sigma ; \{w\})}{\partial w} \\\nonumber
    &= \sum_{\sigma} \Delta E(\sigma) \sum_{i\in\sigma} \frac{\partial \textup{ANN}_{t_i}(\sigma_i^{R_c} ; \{w_{t_i}\})}{\partial w}
    \\\nonumber
    &\textup{where} \quad \Delta E(\sigma) = E^{\textup{ANN}}(\sigma ; \{w\}) - E^{\textup{ref}}(\sigma)
    \quad ,
\end{align}
which can be efficiently calculated using backpropagation.
The computational cost and memory requirement of energy training is, per data point, independent of the size of the data set but instead scales as $\mathcal{O}(N_w)$ where $N_w$ is the number of weight parameters.
Therefore, for training data sets containing a total of $N_{atom}$ atoms, the total computational cost scales linearly with data points $\mathcal{O}(N_w N_{atom})$, which makes this approach feasible for large data sets up to millions of data points.

However, while energy training is computationally efficient, it does not fully utilize reference data, since it discounts the interatomic forces, which provide valuable high-dimensional information about the PES gradient and can be obtained from many electronic structure methods using the Hellmann-Feynman theorem without significant computational overhead~\cite{politzer_hellmann-feynman_2018}.
Consequently, energy training requires larger data sets to sample the PES more finely to accurately reproduce its gradient and curvature, leading to increased computational overhead for electronic structure reference calculations.
Moreover, training exclusively on energies can result in large uncertainties and unreliable force reconstruction, as low energy errors do not necessarily correlate with accurate force predictions.
This issue is further exacerbated by the presence of noise in the energy data, which amplifies force prediction errors as the model overfits to the noise, undermining the reliability of the reconstructed PES~\cite{chmiela_machine_2017}.

\subsection{Direct force training}

The chemical complexity of interface systems might require electronic-structure methods that are computationally demanding, such as hybrid functional DFT calculations, so that an unnecessary excess of reference data for the ANN potential training must be avoided.
Including atomic force information in ANN potential training significantly reduces data requirements by ensuring the training to explicitly encode a physical constraint, the conservation of total energy $\vec{F}_i= - \vec{\nabla}_i E(\sigma)$~\cite{chmiela_machine_2017}.
Incorporating the physics into ANN training not only enables the ANN potentials to accurately reproduce the gradient of PES with less training data but also helps prevent overfitting to noise in the energy data, ensuring more reliable energy and force predictions.

ANN potentials can be trained simultaneously on energies $E^{\textup{ref}}(\sigma)$ and forces $\vec{F}^{\textup{ref}}_{j}(\sigma)$, where the index $j$ is for atoms in structure $\sigma$, by including the force error $\mathcal{L}^{\textup{force}}$ in the total loss function,
\begin{align}\label{eq:force-loss}
    \mathcal{L}^{\textup{total}}
    &= (1 - \alpha) \mathcal{L}^{\textup{energy}} + \alpha \mathcal{L}^{\textup{force}} \\
  	\mathcal{L}^{\textup{force}}
    &= \sum_{\sigma}\frac{1}{2}\Bigr[ \sum_{j \in \sigma}\bigl\{ -\vec{\nabla}_{j} E^{\textup{ANN}}(\sigma ; \{w\}) - \vec{F}^{\textup{ref}}_{j}(\sigma) \bigl\} \Bigr]^2
   \quad ,
\end{align}
where $\alpha$ is a parameter determining the relative contribution of the the force loss $\mathcal{L}^{\textup{force}}$ to the overall loss function $\mathcal{L}^{\textup{total}}$.
We refer to training that minimizes the loss function $\mathcal{L}^{\textup{total}}$, $\{w^{\textup{opt}}\} = \arg\min\{\mathcal{L}^{\textup{total}}\}$, as \emph{direct energy and force training}.

However, direct force training also has a critical drawback: overhead in training cost and memory, since it requires evaluating the derivative $\mathcal{L}^{\textup{total}}$ with respect to $\{w\}$
\begin{align}\label{eq:force-loss-derivative}
    \frac{\partial \mathcal{L}^{\textup{total}}}{\partial w}
    &= (1 - \alpha) \frac{\partial \mathcal{L}^{\textup{energy}}}{\partial w} \\\nonumber
    &- \alpha \sum_{\sigma} \sum_{j \in \sigma} \Delta \vec{F}_{j}(\sigma)
         \sum_{j \in \sigma} \frac{\partial}{\partial w}
                             \vec{\nabla}_{j} E^{\textup{ANN}}(\sigma ; \{w\})
    \\\nonumber
    &\textup{where} \quad  \Delta \vec{F}_{j}(\sigma)
    = -\vec{\nabla}_{j} E^{\textup{ANN}}(\sigma ; \{w\})
        - \vec{F}^{\textup{ref}}_{j}(\sigma)
    \qquad ,
\end{align}
and calculating the term
\begin{align}
    &\sum_{j \in \sigma} \frac{\partial}{\partial w}
      \vec{\nabla}_{j} E^{\textup{ANN}}(\sigma ; \{w\})
  \\\nonumber
    &= \sum_{j \in \sigma} \sum_{i \in \sigma}
        \frac{\partial}{\partial w} \vec{\nabla}_{j}
        \textup{ANN}_{t_i}(\sigma_i^{R_c} ; \{w_{t_i}\})
\end{align}
requires evaluating the second derivative of the neural networks.
As a consequence, the total computational cost of direct force training scales with $\mathcal{O}(N_w N_{\textup{atom}} N_{\textup{local}})$ where $N_w$ is the number of weight parameters, $N_{\textup{atom}}$ is the total number of atoms in the training set, and $N_{\textup{local}}$ is average number of atoms within $2R_c$, where $R_c$ is the cutoff distance of describing the local atomic environment.
The unfavorable quadratic scaling makes direct force training infeasible for complex systems and limits its applications to interfaces essentially with huge data, requiring a more efficient force training method.

\subsection{Representing potential energy surfaces with Gaussian process regression}

Here, we propose using GPR models as surrogate models to efficiently incorporate atomic forces in ANN potential training in an indirect fashion.
Unlike ANNs, which are sometimes referred to as \emph{parametric} models since they are defined by their architecture and weight parameters $\{w\}$, GPRs are \emph{non-parametric} kernel-based ML models, for which the model construction depends solely on the reference data.
For small data sets, full GPR simultaneously fitting to function values and derivatives is the method of choice with respect to accuracy, remarkably reproducing target PESs, and it provides uncertainty without computational overhead, making it possible to further reduce data requirements with active learning~\cite{bartok2010gaussian, bartok_machine_nodate, bartok_machine_2018, bartok2015g}.

The downside is that the computational cost and memory requirements for constructing a full dense GPR model scale as $\mathcal{O}(N^{3})$ and $\mathcal{O}(N^{2})$, respectively, with the training set size $N$~\cite{deringer_gaussian_2021}.
In addition, the cost of inference or prediction for new data points also depends on the size of the reference training data set~\cite{bartok_gaussian_2015}.
The unfavorable scaling can be improved with sparse GPR techniques, but the fundamental dependence on the data size remains.

\subsection{Indirect force training (GPR-ANN)}
\label{sec:workflow-gpr-ann}

The intrinsic pros and cons of \emph{parametric} ANN and \emph{non-parametric} GPR prompted us to consider a way to integrate the advantages of both approaches.
For interface systems, the overall heterogeneous reference data is naturally comprised of data for subsystems, e.g., bulk structures of the involved materials, cluster structures with different numbers of atoms and compositions, and periodic surface slab models (black boxes in \textbf{Figure~\ref{fig:workflow}}).
Each of these homogeneous subsystems can be individually fitted using separate local GPR models (red lines in \textbf{Figure~\ref{fig:workflow}}), enabling more accurate and specialized representations of their respective PESs.
Using the local GPR models, the overall PES can be finely sampled by perturbing the atomic structures in the subsystem data sets and augmenting synthetic data outside the observed regions with GPR-predicted structure-energy data.
Limiting each GPR model to subsets of the total reference data, mitigates its scalability issues with large data sets, simplifies the fit, and facilitates highly efficient inference.

This process also makes it easy to perform active learning iterations:
The GPR uncertainty of each additionally sampled structure ($\sigma^{'}$ in \textbf{Figure~\ref{fig:workflow}}) is evaluated, and when it exceeds a user-defined threshold, the structure is evaluated by the reference electronic-structure method and added to the reference data set.
Finally, ANN potentials (orange line in \textbf{Figure~\ref{fig:workflow}}) are trained on GPR-augmented energies (red triangles) as well as the original electronic-structure reference energies (black circles) within the efficient energy-only training scheme described above.
Since the synthetic data points generated with the local GPR models are based on the energies and atomic forces, the resulting ANN potentials are also implicitly trained on force information, and we refer to this approach as \emph{indirect force training}.

As will be shown in the \emph{Results} section, a multiple of $M$ between 10 and 40 in synthetic data points relative to the original data is sufficient to obtain the saturated optimal GPR-ANN potentials, which leads to a total computational cost that is significantly lower than direct force training, especially for interface systems consisting of a large number of atoms.

In the following, we compare the training performance of the GPR-ANN approach in terms of accuracy, robustness, and computational efficiency against energy-only training, direct force training, and training using a first-order data augmentation method~\cite{cooper_efficient_2020} referred to as Taylor-ANN here.

\subsection{\ce{H2} molecule}

\begin{figure*}[t]
  \centering
  \includegraphics[width=0.95\textwidth]{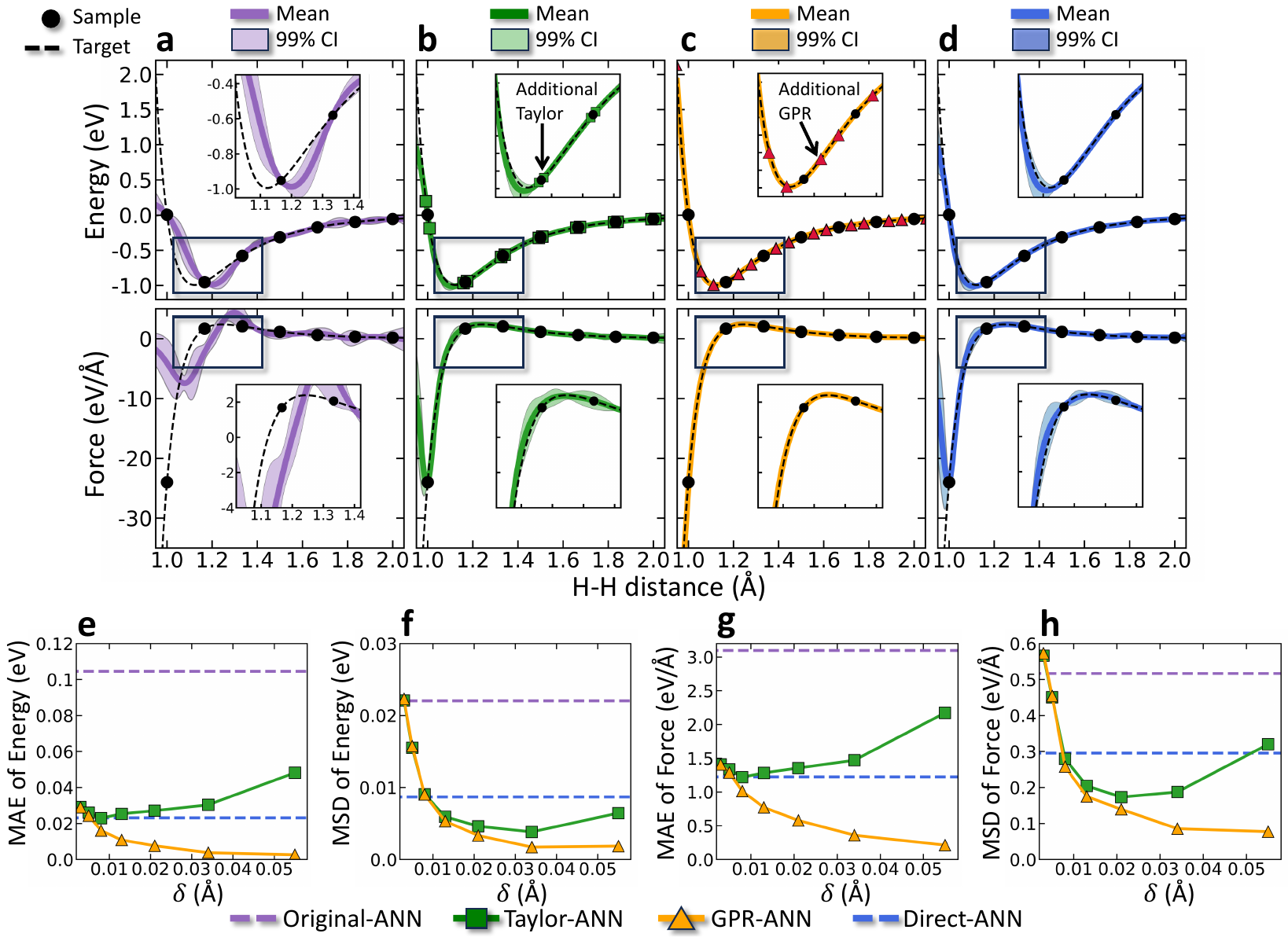}
  \caption{\label{fig:H2-ANN-energy}
  \textbf{Comparison of the different ANN potential training strategies for an \ce{H2} molecule.} The same seven reference data points (black circles) sampled from the target potential energy surface of a \ce{H2} dimer (dashed black line) were used to assess the accuracy and robustness of ANN potentials obtained by training with the four strategies detailed in the main text: \textbf{a}, energy-only training, indirect force training with \textbf{b}, the Taylor-expansion method and, \textbf{c}, the GPR-ANN method, and \textbf{d}, direct force training. The insets show zoomed-in views of the regions marked with rectangles. The mean predicted energies (top) and forces (bottom) of 10~ANN potentials are shown as solid lines, and the shaded regions indicate the 99\% confidence interval (CI) as a measure of uncertainty. For the data-augmentation approaches, Taylor-ANN and GPR-ANN, the seven reference energies were supplemented with 14 predicted energies (green squares in \textbf{b} and red triangles in \textbf{c}), and the corresponding \ce{H2} structures were generated with atomic displacements of $\delta=\pm{}0.008$~Å and $\delta=\pm{}0.055$~Å, respectively. The Taylor-ANN and GPR-ANN potentials corresponding to the optimal atomic displacements are shown, and results from other $\delta$ variables can be found in \textbf{Figures~S4--7}. The accuracy and robustness of the training strategies are quantified by the \textbf{e}, mean absolute error (MAE) and, \textbf{f}, mean standard deviation (MSD) of the energy and the \textbf{g}, MAE and \textbf{g}, MSD of the force, respectively. For the data-augmentation methods, these measures depend on the displacement length and are shown as a function of $\delta$.}
\end{figure*}

As a first example, we consider a Lennard-Jones (LJ) potential roughly approximating the \ce{H-H} dimer for the purpose of an intuitive PES visualization.
In \textbf{Figure~\ref{fig:H2-ANN-energy}}, the target PES in the bond length range from 0.95 to 2.05~Å is displayed as a dashed black line, and seven reference samples are marked by black circles.
The accuracy and robustness of the different ANN training methods are examined in terms of the mean and standard deviation (SD) over a committee of 10~ANN potentials to visualize how accurately the mean reproduces the target PES and how robust the training result is based on the SD evaluating the variation between each of the 10~ANNs.
Results for the predicted energies (top panels) and forces (bottom panels) are shown in \textbf{Figure~\ref{fig:H2-ANN-energy}\textbf{a-d}} for each of the four ANN training methods with the mean and the 99\% confidence interval (CI) evaluated from the SD represented by solid lines and shaded regions, respectively.

In the case of the data-augmentation approaches, Taylor-ANN and GPR-ANN, the seven reference forces are translated into additional energies for 14 synthetic structure-energy data points, and the combined total of $7+14=21$ energy data points were used as training data.
The additional structures were generated by displacing the atoms in the seven original reference structures by small displacements $\delta$, and the energies approximated by linear Taylor expansion (green squares) and a GPR model (red triangles) are shown in the figure as well.

Note that the performance of the Taylor-ANN and GPR-ANN potentials depends on the choice of the displacement length $\delta$, and the results shown in \textbf{Figure~\ref{fig:H2-ANN-energy}b} and \textbf{c} are from the potentials with the optimal displacements $\delta$.
\textbf{Figure~S1} shows the approximate energies in comparison to the reference LJ PES for different displacements $\delta$ ranging from $\pm{}0.003$ to $\pm{}0.055$~Å.
The mean absolute errors (MAE) relative to the LJ reference energies are also summarized in \textbf{Figure~S1d} as a function of the displacement length.
As the displacement length increases, the Taylor-ANN energies deviate farther from the reference due to the limitations of the first-order approximation, while the GPR-ANN energies agree closely with the LJ PES regardless of the displacements considered.
The ANN potentials trained on augmented energy data highly depend on the perturbation size and correctness of corresponding energies that the potentials were trained on, and \textbf{Figures~S2--7} summarize the mean, SD, and error of the mean with respect to target LJ PES.
The best Taylor-ANN potentials were obtained with $\delta=\pm{}0.008$~Å and the best GPR-ANN potentials were obtained with the largest $\delta=\pm{}0.055$~Å.

All of the four training methods exactly reproduce the energies given as training data with negligible uncertainty.
However, the error of mean over 10 ANN potentials obtained from energy-only training increases in between training structures (\textbf{Figure \ref{fig:H2-ANN-energy}a}).
The errors come from an incorrect reproduction of the PES gradient, even at the reference points, as is evident from the force errors shown in the bottom panels of the figures.
In addition, for the smooth PES region at long \ce{H-H} separations, the SD is not negligible despite the mean of the 10 ANN potentials aligning well with the target PES, showing the interpolation instability of energy-only training with insufficient data.

Indirect force training with the Taylor-ANN approach (\textbf{Figure~\ref{fig:H2-ANN-energy}b}) or direct force training (\textbf{Figure~\ref{fig:H2-ANN-energy}d}) corrects the slope of the ANN potentials near the reference samples, leading to great improvements in interpolation as shown in a significant reduction in the error and SD for structures not included in the training data but located nearby.
However, the SD still remains non-negligible, particularly for structures in between the training data as shown in the insets.
Additionally, the mean demonstrates a significant underestimation of absolute forces in the repulsive region where the \ce{H-H} distance is below 1~Å, highlighting the intrinsic extrapolation limitations of ANN training beyond the scope of local training data, small $\delta$ of $\pm{}0.008$~Å in the Taylor-ANN method, and local forces in direct force training.
This issue persists even with the use of large perturbations in the Taylor-ANN approach, as the model learns from inaccurate additional energy values (\textbf{Figure~S4g}~and~\textbf{S6g}).

On the other hand, the GPR-ANN potentials trained on accurate augmented energies for the diverse structures generated with large $\delta$ achieve the most accurate reproduction of the PES and its derivatives, maintaining negligible SD both within and beyond the training data range as shown in \textbf{Figure~\ref{fig:H2-ANN-energy}c}.

The four training methods are further compared in terms of their mean absolute error (MAE) and mean SD (MSD) over 200 test points in \textbf{Figures~\ref{fig:H2-ANN-energy}e--h}.
The results are shown as a function of the displacement length for the Taylor-ANN and GPR-ANN approaches.
Taylor-ANNs (green squares) show the best accuracy for a small displacement of $\delta=0.008$~Å, but their uncertainty is the lowest at a much larger displacement of $\delta=0.034$~Å.
Additional structures generated by small displacements are very similar to the reference structures, and thus, there are still large PES regions that are not well sampled.
In general, if the synthetic data points are too similar to the original data, i.e., if the displacements $\delta$ are chosen too small, the data-augmentation methods Taylor-ANN and GPR-ANN do not show any notable improvement in robustness compared to energy-only training, as seen in \textbf{Figure~\ref{fig:H2-ANN-energy}f} and \textbf{h}.
As the displacement increases, the additional structures are more distinct from the original reference structures, and these well-distributed training data greatly reduce the variance among ANN potentials.
At the same time, however, Taylor-ANN potential energies become less accurate as the displacement increases (\textbf{Figure~S1}), and including the inaccurate synthetic data in the training data degrades the ANN potential accuracy despite decreasing the uncertainty in the predicted energies and forces.
Thus, the Taylor-ANN augmentation method suffers from a trade-off between data diversity and accuracy that needs to be accounted for when it is used.

In contrast, we can see that the GPR-ANN augmentation method is able to provide accurate energy labels for highly displaced unique structures.
As a result, ANN potentials trained on the GPR-augmented energy data set show a gradual improvement in accuracy and uncertainty with increasing displacement length.
The GPR-ANNs with the largest displacement of $\delta=0.055$~Å, which results in the most uniform sampling of the PES regions, almost perfectly represent the LJ PES, exhibiting excellent accuracy and robustness in both interpolation and extrapolation regions.
The MAEs of the GPR-ANN potentials for energies and forces are 3~meV and 0.21~eV/Å, lower than the MAE achieved by energy-only training (105~meV, 3.10~eV/Å) and direct force training (23~meV, 1.22~eV/Å).
In addition, the MSDs of the GPR-ANN potentials for energies and forces are 2~meV and 0.08~eV/Å, i.e., also lower than the MSD for energy-only (22~meV, 0.52~eV/Å) and direct force training (9~meV, 0.30~eV/Å).

Given identical reference energy and force data, the GPR-ANN data-augmentation strategy makes optimal use of the available information and leads to the most accurate and robust (least uncertain) potentials among the four considered ANN training methods.
In practice, this means the GPR-ANN method requires the least number of reference electronic-structure calculations to reach a desirable level of accuracy and uncertainty.

While the dihydrogen molecule is a test system that is easy to conceptualize, it does not reflect the complexity of real-world applications.
Therefore, we next compare the ANN training approaches for a higher-dimensional system comprised of two ethylene carbonate (EC) molecules.

\subsection{Ethylene carbonate molecule dimers}

To assess the GPR-ANN data-augmentation method for a relevant application, we first turned to the electrolyte side of the electrolyte-electrode interface that we seek to model.
The energies and atomic forces of 1,000 ethylene carbonate (EC) dimer structures were evaluated with hybrid-functional DFT calculations, and the resulting data set was divided into 250 training and 750 test data points.
See the \emph{Methods} section for details of the DFT calculations and structure generation.

\begin{figure*}[t]
    \centering
    \includegraphics[width=0.7\textwidth]{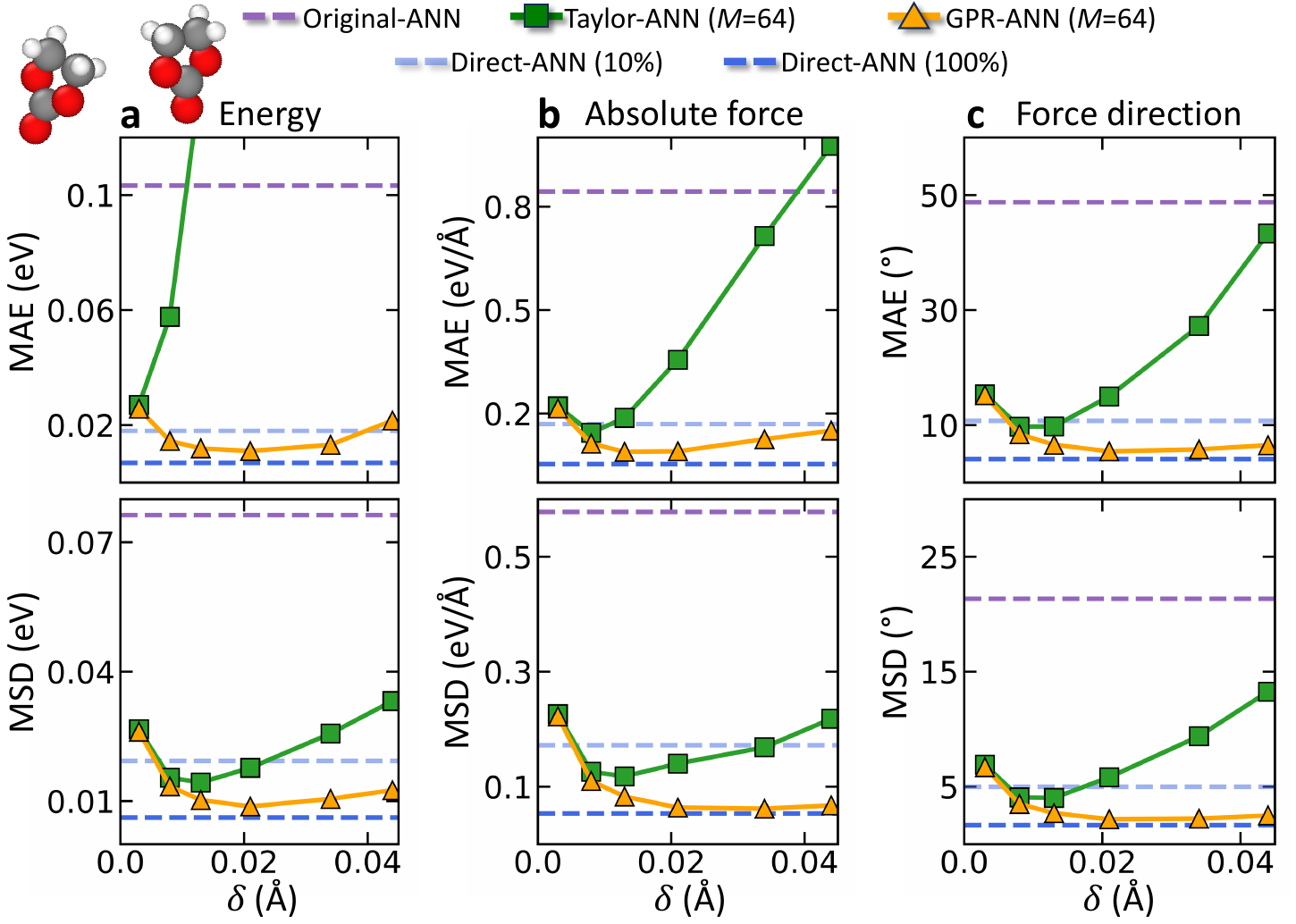}
    \caption{\label{fig:EC-EC-average}
        \textbf{Comparison of the accuracy and robustness of the four ANN training methods for ethylene carbonate dimer structures.} The mean absolute error (MAE) and mean standard deviation (MSD) over a committee of 10 ANN potentials are shown for \textbf{a}~the energy, \textbf{b}~the absolute magnitude of the forces, and \textbf{c}~the force direction. These metrics are shown for ANN potentials obtained from energy-only training (dashed purple line), indirect force training with the Taylor-ANN (green squares), and the GPR-ANN (orange triangles) approach, and direct force training with 10\% forces (dashed light blue line) and 100\% force information (dashed dark blue line).}
\end{figure*}

As for the \ce{H2} example before, we compared the accuracy and robustness of the four different ANN training methods by evaluating the MAE and MSD for the energy and force predictions for the 750~test structures using a committee of 10~ANN potentials.
\textbf{Figures~\ref{fig:EC-EC-average}a-c} show the MAE and MSD of the energy, absolute force, and force direction, respectively, as a function of the displacement amplitude $\delta$.

Each EC, \ce{(CH2O)2CO}, consists of 10 atoms, so for two molecules the total number of force components is 60 for each EC dimer structure.
With direct force training, the percentage of force components to consider during training is a user parameter, and prior research suggests that $\sim$10\% of the force components can already be sufficient and provide an optimal balance of computational efficiency and training outcome~\cite{lopez-zorrilla_aenet-pytorch_2023}.
Therefore, \textbf{Figures~\ref{fig:EC-EC-average}a-c} show the performance metrics of direct force training with 10~and~100\% force information, respectively.

As in the \ce{H2} example, the indirect force training with the Taylor-ANN approach achieves the lowest MAE for $\delta{}=0.003$~Å, which is again a smaller displacement than the one that minimizes the MSD ($\delta{}=0.013$~Å).
Atomic forces predicted by the Taylor-ANN potentials are comparable to those predicted by GPR-ANN potentials for $\delta<0.01$~Å, but the performance of the GPR-ANN potentials remains robust even for larger displacements due to an increase in data diversity with more accurate energy augmentation achieved through GPR models, which better reflect the true PES compared to linear Taylor expansion.
In all considered metrics, accuracy and robustness for energies and forces, the GPR-ANN potentials with $\delta>0.003$~Å are better than ANN potentials from direct force training with 10\% of the force information.
With an optimal $\delta$ of 0.021~Å, GPR-ANN potentials are comparable to direct force training with 100\% forces, exhibiting MAE and MSD values that improve over energy-only training by about one order of magnitude.

Note that the data shown in \textbf{Figures~\ref{fig:EC-EC-average}a-c} is for Taylor-ANN and GPR-ANN potentials with a multiple of $M=64$.
As seen in supplemental \textbf{Figure~S8}, GPR-ANN potentials improve with increasing multiple and converge quickly, outperforming direct force training with 10\% of the force data already at a multiple of $M=16$ and for multiples greater than 36 becoming comparable to direct force training with 100\% forces.

\begin{figure*}[t]
  \centering
  \includegraphics[width=0.7\textwidth]{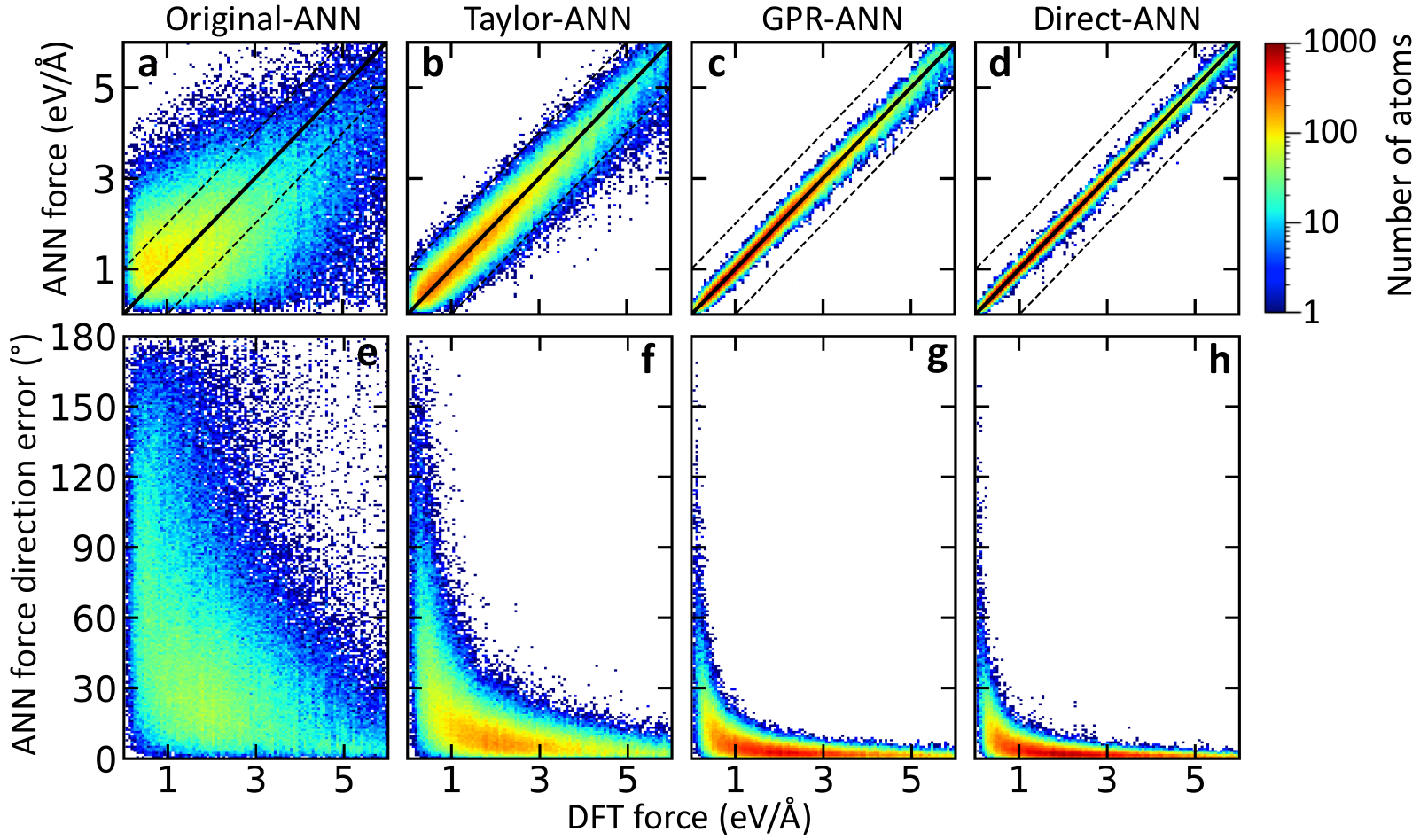}
  \caption{\label{fig:EC-EC-force}
  \textbf{Detailed analysis of the atomic forces in ethylene carbonate dimers predicted with the different training approaches.} \textbf{a-d},~Correlation of the magnitude of the forces predicted by ANN potentials with the DFT reference. \textbf{e-h}, Error in force direction with respect to DFT reference. The predictions were made by a committee of 10 potentials obtained from energy-only training (\textbf{a}, \textbf{e}), implicit force training with the Taylor-ANN (\textbf{b}, \textbf{f}) and GPR-ANN (\textbf{c}, \textbf{g}) methods, and direct force training (\textbf{d}, \textbf{h}). The color indicates the frequency of occurrence using a logarithmic scale. The solid black line in the top panels \textbf{a-d} corresponds to perfect correlation with the DFT reference, and the dashed black lines indicate differences greater than 1~eV/Å. Optimal parameters were used for all force training methods: Taylor-ANN ($\delta$=0.003~Å, multiple=64), GPR-ANN ($\delta$=0.021~Å, multiple=64), and direct force training (100\% forces, alpha=0.3).}
\end{figure*}

\textbf{Figures~\ref{fig:EC-EC-force}a-d} show a more detailed analysis of the atomic forces predicted by ANN potential trained with the different strategies.
The figure shows the correlation of the predicted absolute force magnitude with the DFT reference forces.
ANN potentials trained on the energy only (Original-ANN in \textbf{Figure~\ref{fig:EC-EC-force}a}), clearly do not provide accurate force predictions without expanding the EC dimer database.
For a large fraction of the atoms, the force error is greater than 1~eV/Å, which is the range indicated with thin dashed lines in the figure.
The three other training approaches greatly improve the absolute force distribution, and the figure shows results for optimal parameters: Taylor-ANN potentials ($\delta$=0.003~Å, multiple=64), GPR-ANN potentials ($\delta$=0.021~Å, multiple=64), and direct force training (100\% forces, $\alpha$=0.3).
Potentials trained with the GPR-ANN approach and direct force training show similar performance in force prediction, achieving that the absolute forces of most of the atoms are predicted to be close to the DFT reference, i.e., close to the $x=y$ diagonal of the plots.

\textbf{Figures~\ref{fig:EC-EC-force}e--h} show the corresponding distribution of ANN prediction errors in the direction of the atomic force vectors as a function of the absolute value of the DFT force.
Again, the failure of the ANN potentials trained on 250 energies only is obvious, and the forces acting on a significant fraction of the atoms are predicted in opposite direction (180°) relative to the reference.
The errors in atomic force direction are also significantly reduced when force information is included, especially with the GPR-ANN approach or direct force training.
For these two approaches, errors in force direction larger than 30° only occur in a small fraction of the atoms and for force vectors with small magnitudes below 0.5~eV/Å.

While the performance of the GPR-ANN approach looks promising for the EC molecule example, interface systems are yet more challenging to model.
Therefore, as a final test, we will compare the different training methods for EC adsorbed on and interacting with the surface of Li metal.

\subsection{EC on the surface of lithium metal}

We generated 800 reference structures of an EC molecule on the Li(100) surface, 46 atoms in total, by applying random displacements to all atoms in the ground-state configuration.
All structures were labeled with energies and atomic forces from hybrid-functional DFT calculations, the details of which are provided in the \emph{Methods} section.
The data set was split into 200 training and 600 test data points (see \emph{Methods} section).

\begin{figure*}[t]
    \centering
    \includegraphics[width=0.7\textwidth]{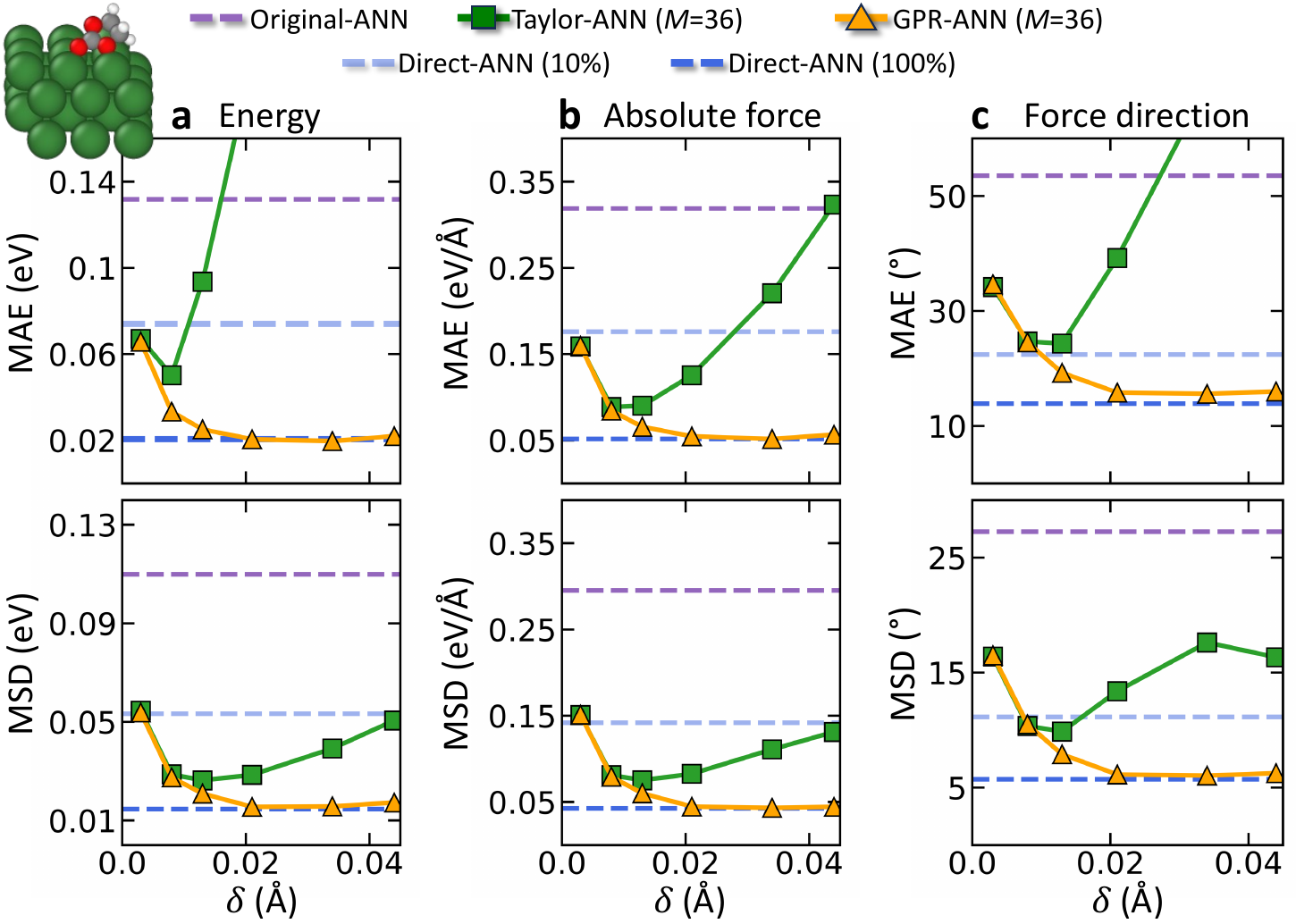}
    \caption{\label{fig:Li-EC-average}
        \textbf{Comparison of the accuracy and robustness of the four ANN training methods for an ethylene carbonate molecule adsorbed on the lithium metal (100) surface.} The mean absolute error (MAE) and mean standard deviation (MSD) based on a committee of 10~ANN potentials are shown for the \textbf{a}, energy, \textbf{b}, absolute force magnitude, and \textbf{c} force direction. Results are shown for energy-only training (dashed purple lines), indirect force training with the Taylor-ANN (green squares) and GPR-ANN (orange triangles) data-augmentation methods, and direct force training with 10\% (dashed light blue lines) and 100\% (dashed dark blue lines) force information.}
\end{figure*}

As for the previous systems, committees of 10 ANN potentials were used to predict the energy and force of the 600 test structures, and the MAE and MSD are summarized in \textbf{Figure~\ref{fig:Li-EC-average}}.
Overall, the trends are similar as for the EC dimer structures from the previous section, yet more pronounced:
Implicit force training with the Taylor-ANN approach achieves the lowest MAE and MSD for a $\delta$ of 0.008~Å, and the accuracy and robustness of the method for energies and the magnitude of the forces lie between those of direct force training with 10\% and 100\% force information and is comparable to direct force training with 10\% for the force direction.
However, the errors and variance rapidly increase with increasing $\delta$ as the first-order Taylor expansion becomes unreliable.
In practice, it can be expected to be challenging to find an optimal $\delta$ that both yields data diversity and provides sufficient accuracy, since the perturbation parameter depends on the unknown PES.

In contrast, the GPR-ANN data-augmentation approach is much more robust with respect to the choice of the displacement amplitude.
As seen in \textbf{Figure~\ref{fig:Li-EC-average}}, all GPR-ANN potentials with $\delta>0.013$~Å show accuracy and robustness comparable to direct force training with 100\% force information across all of the metrics.

The MAE and MSD as a function of the augmentation multiple is plotted in \textbf{Figure~S9} for fixed optimal $\delta$ values of 0.008~Å for the Taylor-ANN and 0.034~Å for the GPR-ANN approach, respectively.
The GPR-ANN approach with a multiple of 16 already reaches the accuracy and robustness of direct force training with 100\% forces for this complex Li-EC system.

\begin{figure*}[t]
  \centering
  \includegraphics[width=0.7\textwidth]{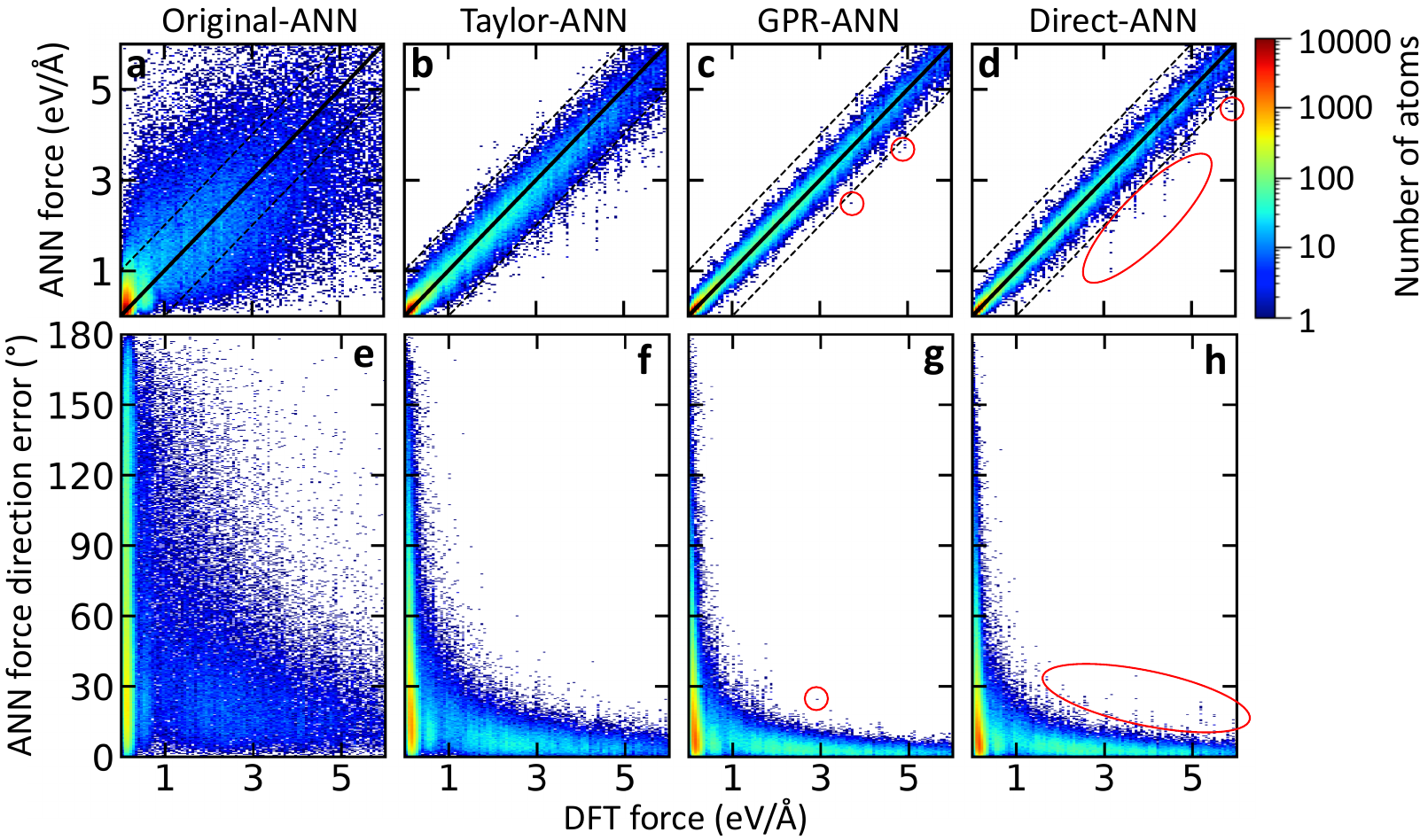}
  \caption{\label{fig:Li-EC-force}
  \textbf{Detailed analysis of the atomic forces in ethylene carbonate adsorbed on the lithium (100) surface predicted by different ANN potentials}. \textbf{a-d}~Correlation between the predicted absolute force and the DFT reference. \textbf{e-h} Error in the force direction with respect to DFT reference. The predictions are based on a committee of 10 ANN potentials obtained from energy-only training (\textbf{a}, \textbf{e}), indirect force training with the Taylor-ANN (\textbf{b}, \textbf{f}) and GPR-ANN (\textbf{c}, \textbf{g}) methods, and direct force training with 100\% force information (\textbf{d}, \textbf{h}). The color encodes the frequency of occurrence with a logarithmic scale. The solid black line in the top panels \textbf{a-d} corresponds to a perfect agreement with the DFT reference, and the dashed black lines indicate deviations greater than 1~eV/Å. For all force training methods, optimal parameters were used: Taylor-ANN ($\delta$=0.008~Å, multiple=36), GPR-ANN ($\delta$=0.034~Å, multiple=36), and direct force training (100\% forces, alpha=0.3).}
\end{figure*}

\textbf{Figure~\ref{fig:Li-EC-force}} shows a detailed analysis of the atomic forces with correlation plots and the directional errors, in the same fashion as above in \textbf{Figure~\ref{fig:EC-EC-force}} for the EC dimer case.
The trends are similar to those seen for the EC dimers, and energy-only training on 200 energy data points proves certainly unreliable regarding the predictions for both magnitude and direction of forces showing severe errors for a large fraction of atoms.
The best agreement with the DFT reference was achieved by the GPR-ANN potentials and direct force training, consistent with the average analysis of \textbf{Figure~\ref{fig:Li-EC-average}}.
For this more challenging system, direct force training shows significantly more outliers with errors greater than 1~eV/Å (highlighted in red circles) than training with the GPR-ANN approach, implying that direct force training is more vulnerable to critically large errors despite the comparable MAE and MSD of the two methods.

\section{Computational efficiency of the GPR-ANN method}
\label{sec:efficiency}

As discussed above, the computational cost of direct force training scales with $\mathcal{O}(N_w N_{\textup{atom}} N_{\textup{local}})$ and that of synthetic data GPR-ANN training with $\mathcal{O}(N_w N_{\textup{atom}} M)$.
As the examples of the previous sections demonstrated, a fixed multiple of $M=10$~to~$40$ additional structures generated via random atomic displacement is sufficient to obtain GPR-ANN potentials comparable to direct force training with 100\% force information in regards to the accuracy and robustness for all considered metrics.
For condensed phases and typical cutoff radii, the number of atoms within the local atomic environment, $N_{\textup{local}}$, is at least one order of magnitude greater than $M$ and can be substantially larger for materials with high density.
On the other hand, the fitting of the GRP models also requires computation, which gives rise to a pre-factor.
Therefore, in the following, we benchmark the efficiency of the indirect force training approach by comparing the memory and computer time required by GPR-ANN training and direct force training.

For this benchmark, we used the entire reference data set of 5,168 Li-EC DFT interface calculations comprised of 17 heterogeneous subsets.
This includes the above example of a single EC molecule adsorbed on the Li(100) surface (subset 17), and the other subsets consist of different numbers of atoms and compositions generated with different protocols to sample diverse structural configurations of the interface (see the methods section for details).
Each of the subsets was divided into training and test data, and the total numbers of reference training and test points are 2,100 and 3,068, respectively (see the \emph{Methods} section for details).

\begin{figure}[t]
  \centering
  \includegraphics[width=0.4\textwidth]{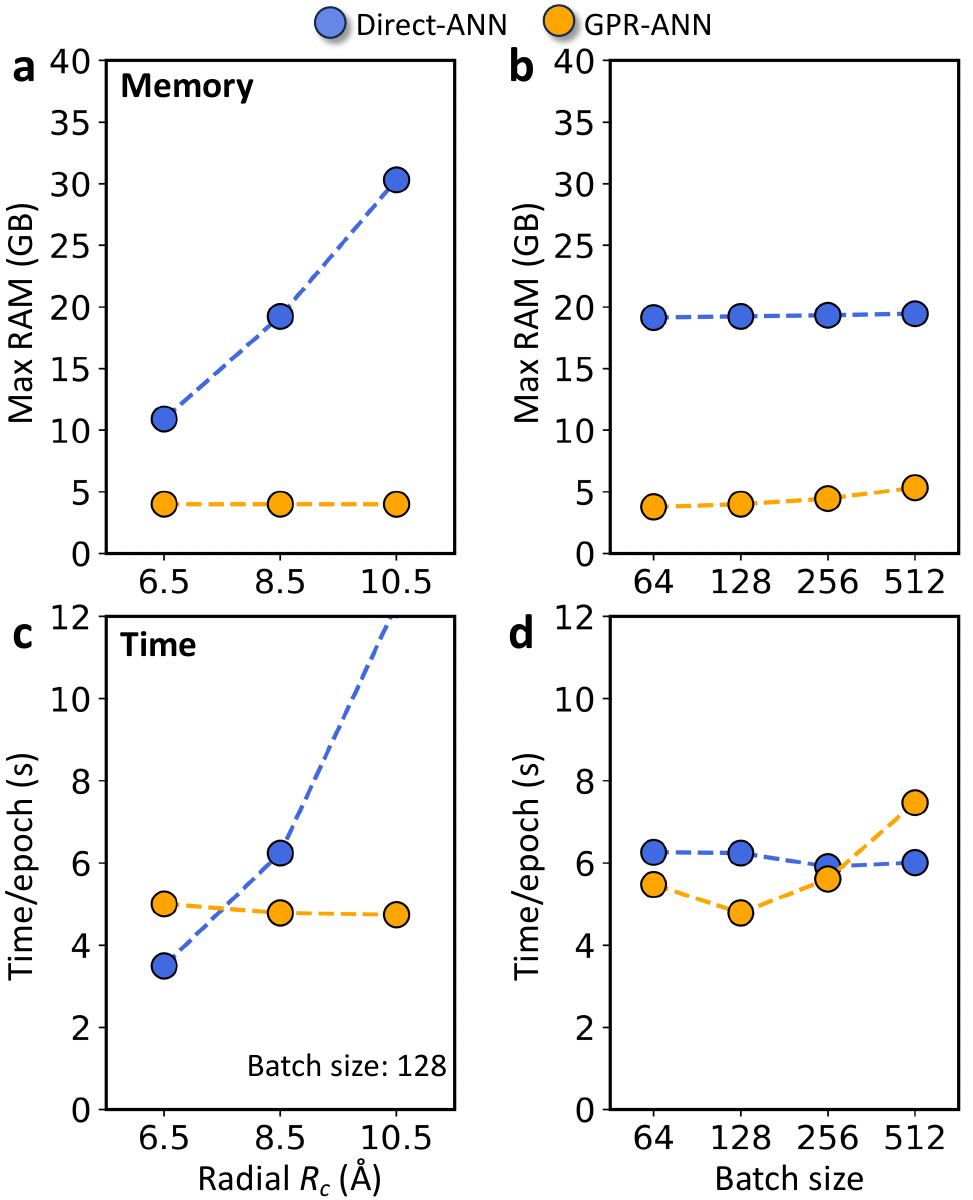}
  \caption{\label{fig:time-info}
  \textbf{Comparison of the memory and computer time required by GPR-ANN and direct force training for Li-EC interface structures.} Maximum random-access memory (RAM) required for direct force training (blue circles) and GPR-ANN training (orange circles) as a function of \textbf{a} the radial cutoff radius and \textbf{b} the batch size. The training times per epoch for both training methods are shown as a function of \textbf{c} the radial cutoff radius and \textbf{d} the batch size.}
\end{figure}

\textbf{Figure~\ref{fig:time-info}a--d} compares the efficiency of direct force training and GPR-ANN training in terms of memory usage and training time per epoch across various choices of batch sizes and cutoff radii (${R_c}$) for the atomic environment descriptors ($\sigma_i^{R_c}$ in Equation~\ref{eq:ANN-energy}).
The ANN potentials were trained on a single CPU of our local computer cluster (Intel Xeon Gold 6226 2.9 GHz).
Both methods were trained using the optimal parameters identified in the previous Li-EC example, i.e., $\delta=0.034$ Å and $M=36$ for the GPR-ANN approach and 100\% forces and $\alpha{}=0.3$ for direct force training.

The GPR-ANN approach benefits from the trivial parallelism of the local GPR models, which can be fitted separately for each data subset (\textbf{Figure~\ref{fig:workflow}}).
Furthermore, fitting GPR models on small homogeneous data sets containing 50--150 structures and generating additional structures within the local structural spaces contributes negligibly to the overall computational cost, especially in terms of memory use.
Therefore, for relevant cutoff radii and batch sizes, the GPR-ANN training consistently requires less memory than direct force training (\textbf{Figure~\ref{fig:time-info}a--b}), and the computer time is lower or comparable (\textbf{Figure~\ref{fig:time-info}c--d}).
In addition, memory usage and training time with the GPR-ANN approach are essentially independent of the cutoff radius, whereas the computational cost of direct force training grows quadratically with the cutoff for condensed phases.
Despite its lower memory usage compared to direct force training, the GPR-ANN method achieves comparable accuracy in both energy and force predictions, with over an order of magnitude improvement in energy and nearly two orders of magnitude in force predictions compared to energy-only training, as demonstrated in \textbf{Figures~S10 and S11}.

\section{Discussion}
\label{sec:discussion}

Data augmentation has previously been proposed as an approach for implicit force training, and we compared the GPR-ANN method with the Taylor-ANN method by Cooper et al.~\cite{cooper_efficient_2020} that is based on a first-order Taylor expansion.
As seen in the benchmark results, the GPR-ANN approach is significantly more robust with respect to the choice of the additional structures that are labeled with synthetic energies.
Specifically, when additional structures are derived from reference structures via the random displacement of atoms, the Taylor-ANN approach works best for small displacement amplitudes where the potential energy varies approximately linearly with respect to the reference energy.
In contrast, the non-linear GPR models are also able to fit the PES in regions further away from the reference structures.
This is important since the optimal displacement amplitude for the ANN-Taylor method depends on the curvature of the PES and is, therefore, system-dependent.
For example, the optimal displacement amplitude for the EC dimers and the adsorbed EC molecule was 0.003~Å and 0.008~Å, respectively.
Soft bonds (e.g., Li-Li bonds in Li metal) can tolerate greater displacements than stiff bonds (e.g., C-C bonds in EC molecules), and in complex systems such as interfaces, it can become challenging to select displacement amplitudes in practice.
The GPR-ANN approach mostly avoids this parameter dependence and works well for a wide range of displacement amplitudes in our test systems.

GPR models excel at reproducing unknown PESs based on small reference datasets.
Additionally, GPR models provide an intrinsic model uncertainty that can be used to confirm whether the predicted energy is robust, ensuring the diversity and reliability of augmented energy data.
Hence, the GPR-ANN approach can provide benefits not only for efficiently sampling reference training data but also for preventing the inclusion of inaccurate synthetic energy data (\textbf{Figure~\ref{fig:workflow}}).

While the primary purpose of the GPR-ANN method is indirect force training, its model uncertainty also enables efficient active learning, reducing overall data requirements.
Estimating uncertainties with ANNs alone requires computationally expensive committees of multiple ANNs, and thus, the computational effort of training uncertainty models is a multiple of direct force training.
In contrast, in the GPR-ANN training process, GPR surrogate models provide uncertainty estimates at no additional cost, allowing for non-redundant reference data sampling with Bayesian learning strategies prior to training multiple ANNs.
Note that the DFT reference data for the present work were obtained using traditional sampling methods, i.e., molecular dynamics simulations, random atomic displacements, and conformal sampling, and the entire dataset was randomly split into training and test sets (\textbf{Figures~S12, S14, and S16}).

Finally, we conclude that indirect force training is not always the best option.
For low-density materials or molecular data sets, the memory requirements for direct force training can be moderate so that the additional pre-factor of $\sim{}40$ due to additional synthetic data is less favorable than direct force training.
Hence, it depends on the target system and the potential cutoff whether the GPR-ANN approach is effective, and its utility is greatest for condensed-matter systems.

\section{Conclusions}

We introduced a GPR-based data-augmentation approach that indirectly incorporates atomic forces into the training of ANN potentials via synthetic energy data.
The approach bypasses directly training ANN potentials on interatomic forces, which is computationally demanding and can become infeasible due to the quadratic scaling with the range of the potential.
For four test systems with increasing complexity, the dihydrogen molecule, ethylene carbonate dimers, an ethylene carbonate molecule adsorbed on the surface of lithium metal, and heterogeneous data for diverse Li-EC interfaces, we showed that the GPR-ANN approach yields ANN potentials with accuracy and robustness on a par with direct force training across various metrics.
We showed that scaling challenges of the GPR models can be avoided using separate local GPR models, each trained on small subsets of the overall data.
For training on hybrid-functional DFT data of the Li-EC interface system, indirect force training with the GPR-ANN approach significantly lowers the memory requirement compared to traditional direct force training without compromising the training time, ANN potential accuracy, robustness, or transferability.
The GPR-ANN approach, furthermore, provides a model uncertainty without a need for ensemble models that can be used for Bayesian active learning strategies.
As system complexity grows, the ANN-GPR method provides a scalable alternative to traditional direct force training in developing accurate potentials and reduces the need for costly additional reference calculations.
This paves the way for constructing ANN potentials for complex condensed matter systems, such as the interfaces in lithium-ion and lithium-metal batteries.

\section{Methods}
\label{sec:methods}

\subsection{Reference data}

\subsubsection{\ce{H2} molecule}

A Lennard--Jones potential, as implemented in the atomic simulation environment (ASE)~\cite{hjorth_larsen_atomic_2017} library, was used to generate an approximate PES for the \ce{H2} molecule.
The energy and force for $7$ equally-spaced \ce{H-H} bond lengths between $1$~and~$2$~Å were generated as training reference data, and $200$ equally-spaced points between $0.95$ and $2.05$~Å were generated for testing.

\subsubsection{Ethylene carbonate molecule dimers}

1,000 EC dimer reference structures were generated by random displacement of the ground-state structure.
The energies and interatomic forces were evaluated with hybrid-functional DFT calculations using the all-electron electronic structure program FHI-aims in which the Kohn-Sham states are expanded as linear combinations of numerical atomic orbitals~\cite{blum_ab_2009, havu_efficient_2009}.
The HSE06 functional~\cite{heyd_hybrid_2003, heyd_erratum_2006} and FHI-aims' default \emph{tight} basis set were employed for the non-periodic EC dimer structures.
Relativistic effects were taken into account on the level of the zeroth Order Regular Approximation (ZORA)~\cite{van_lenthe_relativistic_1994}.

The 1,000 reference data points were divided into 250 training and 750 test data points, and their relative energy distribution with respect to the minimum ground-state energy is shown in \textbf{Figure~S12}.
Additional structures for implicit force training with the Taylor-ANN and GPR-ANN methods were generated by randomly displacing all the constituent atoms of the 250 training structures.
Random displacements were obtained from a Gaussian distribution, and the amount of displacement was controlled via the standard deviation parameter $\delta$.
\textbf{Figure~S13a} shows the relative energy distribution of Taylor-augmented (green bar) and GPR-augmented (red bar) energies of the same additional structures with different $\delta$ parameters ranging from 0.003~Å to 0.044~Å.
For the GPR-ANN approach, full GPR models were constructed using all the energy and force information of the 250 training structures to predict the energies of the additional structures.

The representation plot shows the predicted energies of 1,000 additional (synthetic) structures as an example, which is $4$ times the number of the reference training structures, i.e., the augmentation multiple is $M=4$.
For $\delta\geq{}0.013$~Å, some of the energies predicted with linear Taylor expansion are lower than the ground-state energy, i.e., have values below zero.
This shows the limitation of the linear Taylor-ANN method since the energy of no structure can be lower than the ground-state energy.

\textbf{Figure S13b} shows the distribution of energy difference between the Taylor- and GPR-augmented energies.
When $\delta$ is small, the Taylor and GPR energies are almost identical, as expected.
However, as $\delta$ increases, the Taylor-expansion energies become increasingly lower than the GPR-augmented energies.
This is the case when the curvature of relevant PES is positive, as schematically described in \textbf{Figure S13c}.

\subsubsection{EC on the surface of lithium metal}

The 800 reference structures were sampled by random displacement of the ground-state structure, and they were evaluated using HSE06 DFT calculations using FHI-aims.
All details of the calculation were the same as in the previous section, except that the Li-EC structures were represented as periodic slab models, and the DFT calculations were performed with $5\times{}5\times{}1$ k-point meshes.
It should be noted that atomic forces obtained from DFT calculations are very sensitive with respect to the density of the k-point meshes, and $5\times{}5\times{}1$ meshes gave converged results in our tests.
\textbf{Figure~S14} shows the relative energy distribution of the reference structures, which were divided into 200 training and 600 test data points.

Additional structures for the data-augmentation approaches were generated by random displacements, as described above for EC dimers, using the same standard deviation parameter $\delta$ as in the previous section.
\textbf{Figure S15a} shows the relative distribution of Taylor-augmented (green bar) and GPR-augmented (red bar) energies of the same synthetic structures corresponding to different $\delta$ values ranging from 0.003~Å to 0.044~Å.
As for the EC dimer, the first-order Taylor expansion underestimates the energy of perturbed structures for large $\delta$ values compared to the GPR energies (\textbf{Figure S15b}).
However, there are some structures 0.008~Å $\leq{}\delta{}\leq{}$ 0.034~Å where the Taylor-expansion energies are higher than the GPR energies, implying that the PES of the Li-EC system is more complex than that of the EC dimers, and reference samples around regions with negative curvature are included as well.

\subsubsection{Heterogeneous data subsets for diverse Li-EC interface structures}

Within non-periodic cluster and periodic slab models, diverse reference structures with different numbers of EC molecules and Li atoms were generated using \emph{ab initio} molecular dynamics (AIMD) simulations, molecular dynamics simulations using preliminary ANN potentials, random displacement, and internal (i.e., conformer) sampling.
All the generated structures were evaluated with HSE06 DFT calculations using FHI-aims with 5$\times$5$\times$1 k-point meshes for periodic cells and a single $\Gamma$ k-point for non-periodic cells.
For each data subset, a representative atomic structure is shown in \textbf{Figure~S16} along with each of the sampling methods and the number of training and test data points.

\subsection{GPR surrogate model}

All GPR construction and GPR-based data augmentation were performed using the ænet-GPR package developed for the present work and available at \url{https://github.com/atomisticnet/aenet-gpr}.
In all examples presented here, full GPR models were utilized, accounting for the covariance between two function values, between a function value and a derivative, and between two derivatives~\cite{bartok_gaussian_2015, chmiela_machine_2017, deringer_gaussian_2021} with the squared exponential as the kernel function.
System-specific parameters are detailed below.

\subsubsection{\ce{H2} molecule}

A GPR model for \ce{H2} molecule was constructed based on the energies and forces of 7 equally-spaced training points.
We used flattened Cartesian coordinates as the global fingerprint.
Using PyTorch's \emph{autograd} functionality, the weight and scale parameters of the kernel function were optimized by iteratively minimizing the energy loss function for the 200 test points.
After 100 iterations, the default parameters of the weight and scale converged to 8.5 and 0.2, respectively (\textbf{Figure~S17}).
\textbf{Figure~17a,b} show the GPR energy and force predictions with the default kernel parameters while \textbf{Figure S17c,d} show the predictions after the parameter optimization.
This GPR model with optimized hyperparameters was used to augment energy data for the GPR-ANN training.

\subsubsection{Ethylene carbonate molecule dimers}

As for the \ce{H2} molecule, flattened Cartesian coordinates were used as structural fingerprints.
We performed a grid search to optimize the GPR kernel parameters for this high-dimensional system, and the optimized parameters, which minimize the energy loss function for the 750 test data points, are 1.0 and 1.5 for the weight and scale, respectively.
\textbf{Figure~S18a,b} show the correlation between the GPR-predicted absolute force and the DFT reference before and after the kernel parameter optimization.
The GPR model fitting to the whole energy and force information of 250 reference training data points with the optimized kernel parameters was used to evaluate GPR-augmented energies.

\subsubsection{EC on the surface of lithium metal}

As fingerprint for the GPR construction for the interface system, we tested both flattened Cartesian coordinates and smooth overlap of atomic positions (SOAP) descriptors~\cite{bartok_representing_2013} as implemented in the DScribe library~\cite{himanen_dscribe_2020}.
A SOAP descriptor was generated with a cutoff radius of 5.0~Å, 6 radial basis functions, and a maximum degree of spherical harmonics of 4.
As summarized in \textbf{Figure~S19}, the GPR model based on the SOAP descriptor (kernel parameters: weight$=$5.0, scale$=$6.0) optimized by the grid search shows an optimal correlation with the DFT reference energies, and this model was adopted as a surrogate model to augment energy data.

\subsubsection{Heterogeneous data subsets for diverse Li-EC interfaces}

For the heterogeneous data set, the descriptor and kernel parameters of the GPR model were not further optimized, and we used the parameters identified as optimal in the previous section.
Using the same kernel parameters, 17 local GPR models were individually fitted to homogeneous training data in each data subset, and the separate local GPR models representing respective PESs were used to generate local synthetic energies for each subsystem.

\subsection{ANN potential training}

All of the ANN training and prediction was carried out using the atomic energy network (ænet)~\cite{artrith_implementation_2016} and ænet-PyTorch~\cite{lopez-zorrilla_aenet-pytorch_2023} packages.
The \emph{Adamw} optimization algorithm with a learning rate of 0.0001 and a regularization parameter of 0.001 was used for all of the training runs.
Atomic environments were represented using a Chebyshev descriptor~\cite{artrith_efficient_2017}.
The system-specific parameters of the Chebyshev descriptors and ANN architectures are described in the following section.

\subsubsection{\ce{H2} molecule}

For the \ce{H-H} dimer, a Chebyshev descriptor was constructed with a radial cutoff radius of 8.0~Å and a radial expansion order of 10.
No angular expansion was used for this linear molecule.
The ANN architecture was $N$-5-5-1, where $N$ is the descriptor dimension, the ANN gives a single output value (the atomic energy), and the two hidden layers each had five nodes.
Hyperbolic tangent activation functions were used.

\subsubsection{Ethylene carbonate molecule dimers}

For the EC-EC dimers, the Chebyshev descriptors the elements \ce{C}, \ce{H}, and \ce{O} were constructed as follows: the radial and angular expansion orders were 12 and 4, respectively, and the radial and angular cutoff radii were 6.5 and 4.0~Å, respectively.
The ANN architecture for each of the elements was 36-10-10-1 with hyperbolic tangent activation functions.
The batch size was 32 for energy-only and direct force training, while a batch size of 256 was used for Taylor-ANN and GPR-ANN training.

\subsubsection{EC on the surface of lithium metal}

For the Li-EC structures, the Chebyshev descriptors used the same parameters as those of the EC molecule above for all elements, and the ANN architecture was also identical.

\subsubsection{Heterogeneous data subsets for diverse Li-EC interfaces}

For the heterogeneous Li-EC database, the same radial and angular expansion orders (12 and 4) were used for the Chebyshev descriptors.
In order to compare the memory and cost overhead with respect to the cutoff radius for atomic descriptors, several different radial and angular cutoff radii were tested: 6.5 and 4.0~Å, 8.5 and 6.0~Å, and 10.5 and 8.0~Å.
The ANN architecture for each of the elements was 36-10-10-1 with hyperbolic tangent activation functions as before, and different batch sizes were tested as described in the main text.


\section{Acknowledgements}

The authors acknowledge support by the Columbia Center for Computational Electrochemistry (CCCE) and computing resources from Columbia University's Shared Research Computing Facility.
J.L.Z. and N.A. thank the Project HPC-EUROPA3 (Grant No. INFRAIA-2016-1-730897) for its support, provided through the EC Research and Innovation Action under the H2020 Programme.

\section*{Data availability}

The authors declare no competing financial interest.
This work made use of the free and open-source atomic energy network (ænet), ænet-PyTorch package.
The source code can be obtained either from the ænet Web site (http://ann.atomistic.net) or from GitHub (\url{https://github.com/atomisticnet/aenet-PyTorch}).
The MD simulations with MLPs input and output files can also be obtained from the GitHub (\url{https://github.com/atomisticnet/XXX}).
The reference \ce{Li/C/H/O} dataset can be obtained from the Materials Cloud repository
(\url{https://doi.org/10.24435/materialscloud:dx-ct}).
The data set contains atomic structures and interatomic forces in the XCrySDen structure format (XSF), and total energies are included as additional meta information.

\section*{Code availability}

\url{https://github.com/atomisticnet/aenet-gpr}

\bibliographystyle{unsrt}
\bibliography{main}


\clearpage\newpage
\onecolumngrid
\appendix
\renewcommand{\thefigure}{S\arabic{figure}}
\renewcommand{\thetable}{S\arabic{table}}
\setcounter{figure}{0}
\setcounter{table}{0}

\section{Supplementary Information}

\section{Supplementary Figures}

\begin{figure}[H]
  \centering
  \includegraphics[width=0.8\textwidth]{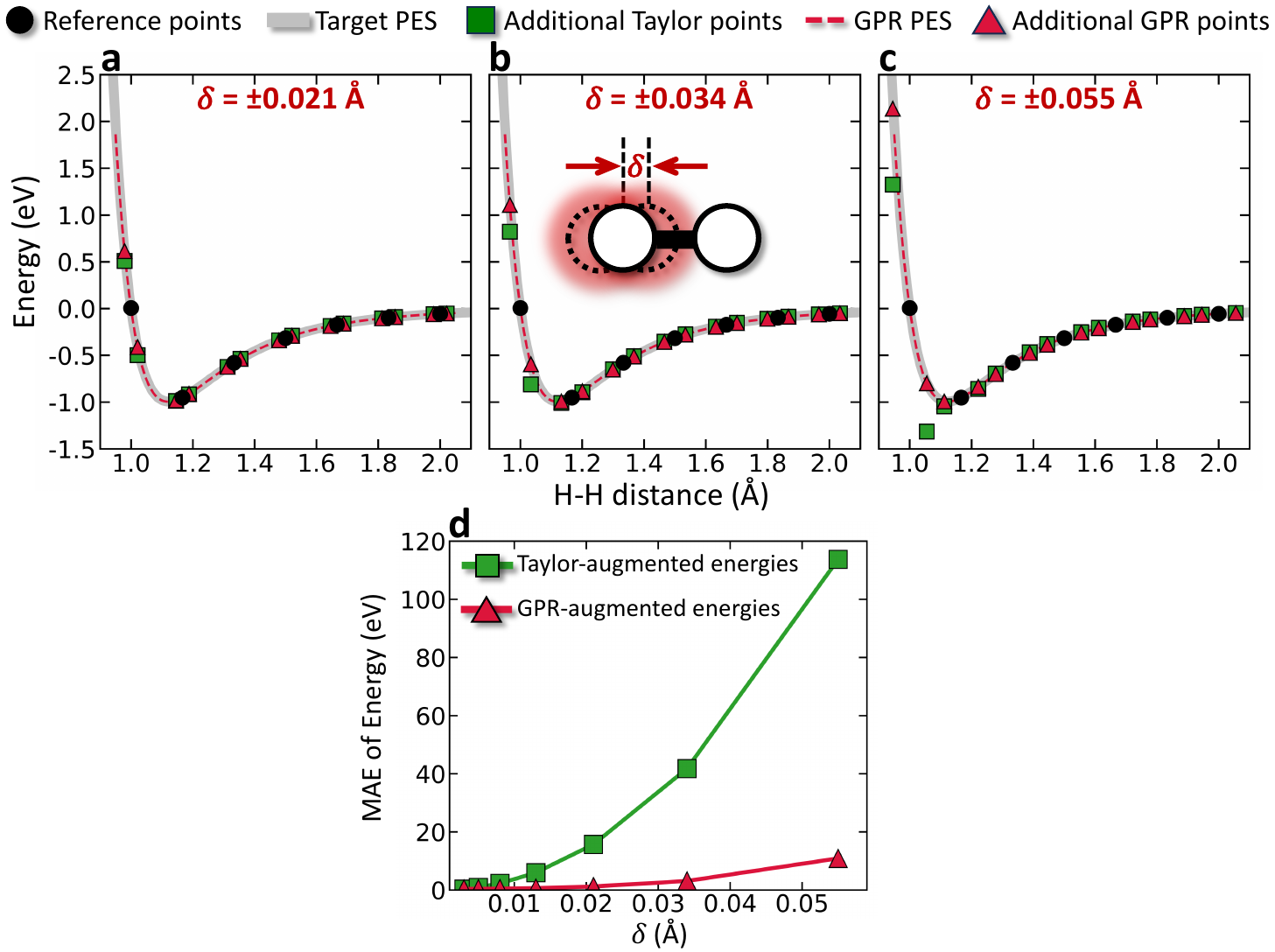}
  \caption{
  \textbf{Synthetic energy data depending on the displacement length and interpolation methods.} The seven reference points (black circles) sampled from the target Lennard-Jones potential of a \ce{H2} dimer (thick gray line) were used to augment energy data. A GPR model (dashed red line) was fitted to the reference points and their slopes. The energies approximated by the linear Taylor expansion (green squares) and GPR model (red triangles) for the additional structures generated with different displacement amplitudes ($\delta$) of \textbf{a} $\pm{}0.021$~Å, \textbf{b} $\pm{}0.034$~Å, and \textbf{c} $\pm{}0.055$~Å. \textbf{d} Comparison of the mean absolute error (MAE) of the Talyor- and GPR-augmented energies as a function of $\delta$.}
\end{figure}

\begin{figure}[H]
  \centering
  \includegraphics[width=0.8\textwidth]{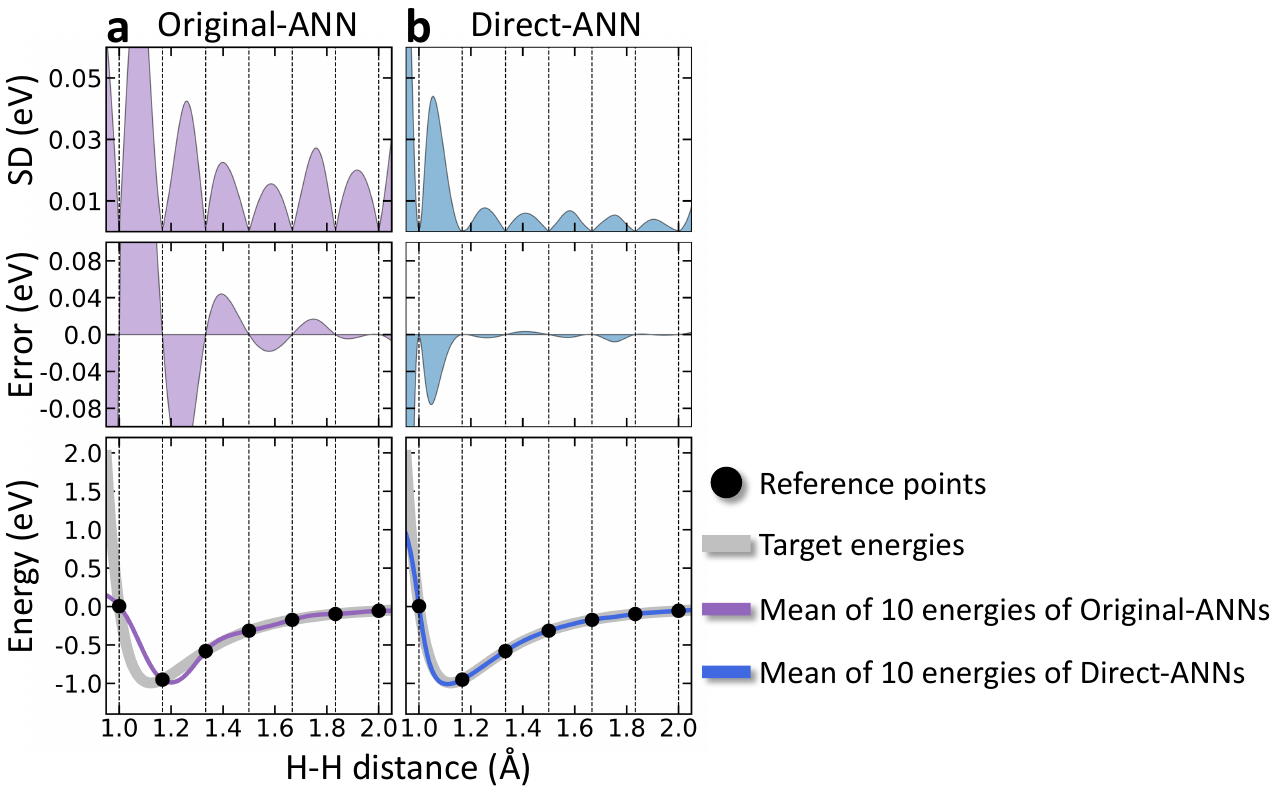}
  \caption{
  Standard deviation (SD) and error of energy predictions over a committee of 10 ANN potentials obtained from \textbf{a} energy training and \textbf{b} direct force training on the seven reference points (black circles). On the bottom panels, the target Lennard-Jones potential energies were represented by thick gray line along with the mean over 10 ANN predictions (solid lines). The error is defined as the difference between the mean and the target potential energy.}
\end{figure}

\begin{figure}[H]
  \centering
  \includegraphics[width=0.8\textwidth]{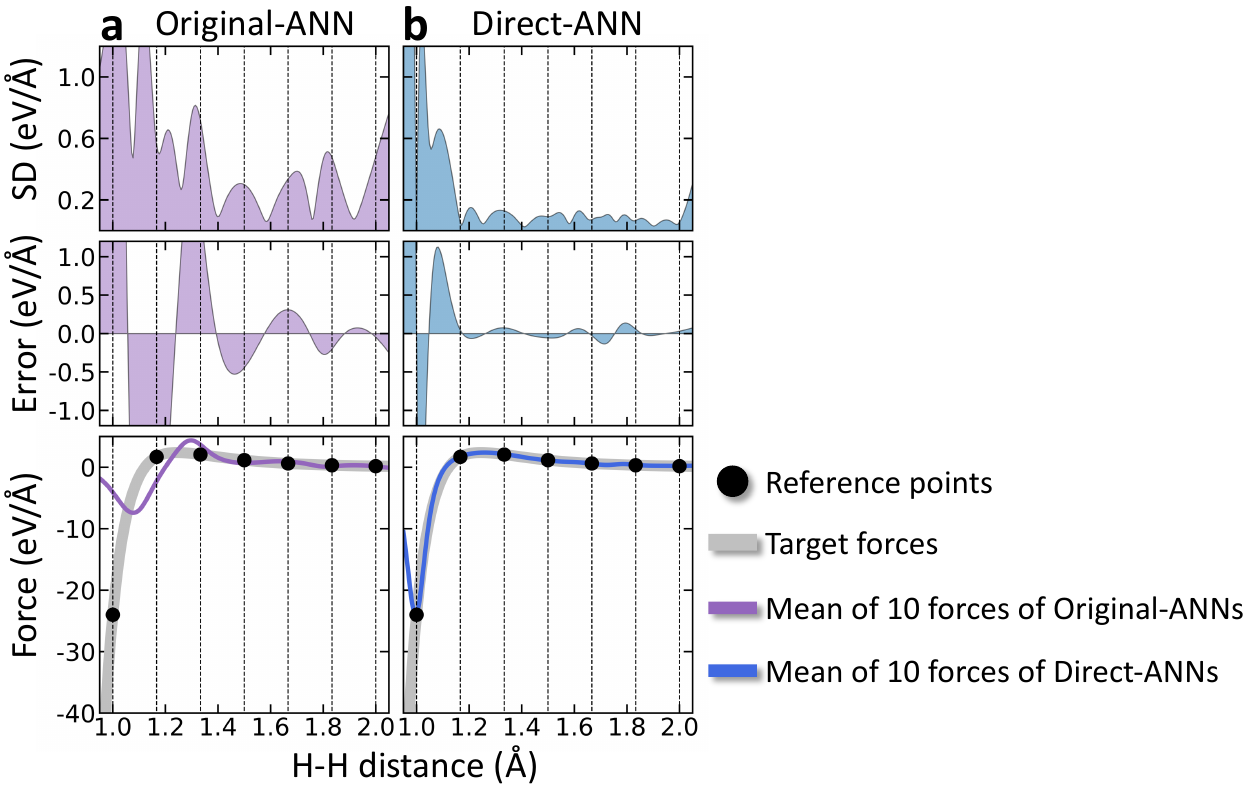}
  \caption{
  Standard deviation (SD) and error of force predictions over a committee of 10 ANN potentials obtained from \textbf{a} energy training and \textbf{b} direct force training on the seven reference points (black circles). On the bottom panels, the target forces, the negative gradient of the Lennard-Jones potential, were represented by thick gray line along with the mean predicted forces (solid lines). The error is defined as the difference between the mean and the target force.}
\end{figure}

\begin{figure}[H]
  \centering
  \includegraphics[width=1.0\textwidth]{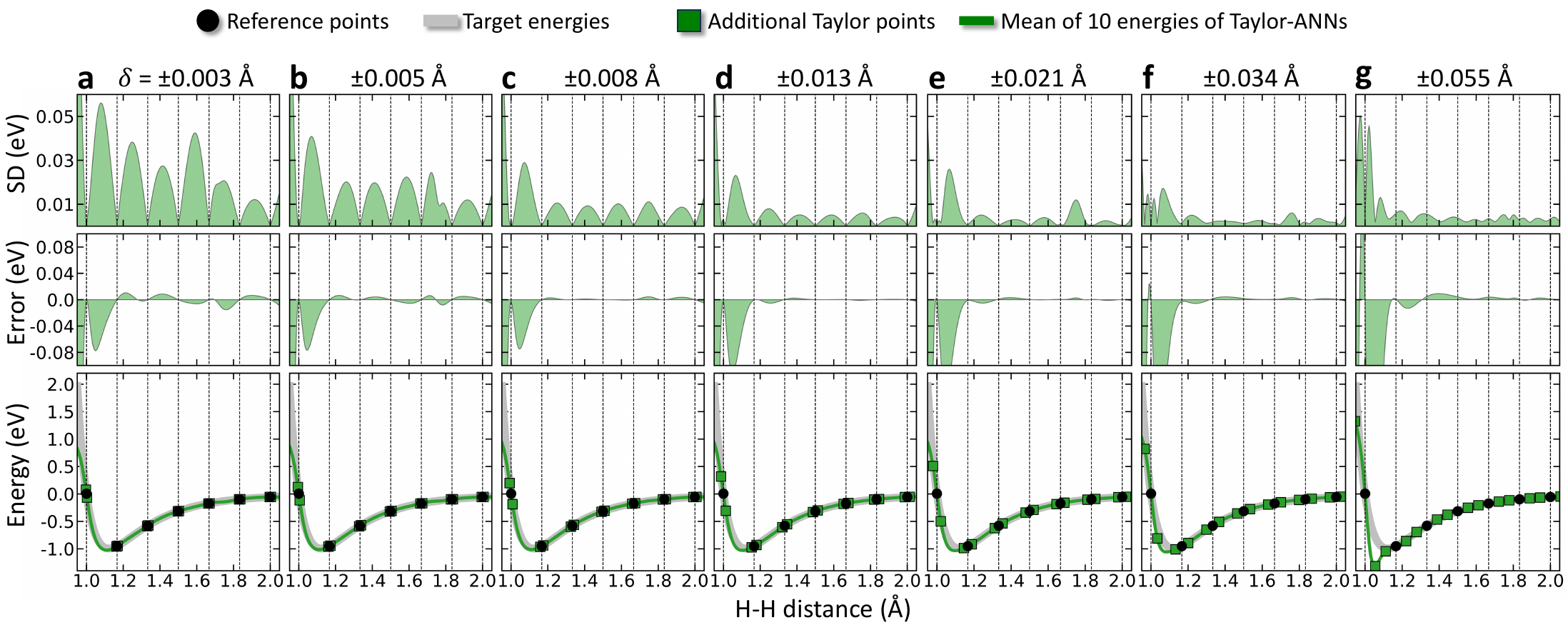}
  \caption{
  Standard deviation (SD) and error of energy predictions over a committee of 10 ANN potentials obtained from indirect force training with the Taylor-expansion method. The potentials were trained on the seven reference points (black circles) and 14 Taylor-augmented energies (green squares) with different displacement amplitudes: \textbf{a} $\pm{}0.003$~Å, \textbf{b} $\pm{}0.005$~Å, \textbf{c} $\pm{}0.008$~Å, \textbf{d} $\pm{}0.013$~Å, \textbf{e} $\pm{}0.021$~Å, \textbf{f} $\pm{}0.034$~Å, and \textbf{g} $\pm{}0.055$~Å. The error is defined as the difference between the mean predicted energies (solid green line) and the target potential energies (thick gray line) shown on the bottom panels.}
\end{figure}

\begin{figure}[H]
  \centering
  \includegraphics[width=1.0\textwidth]{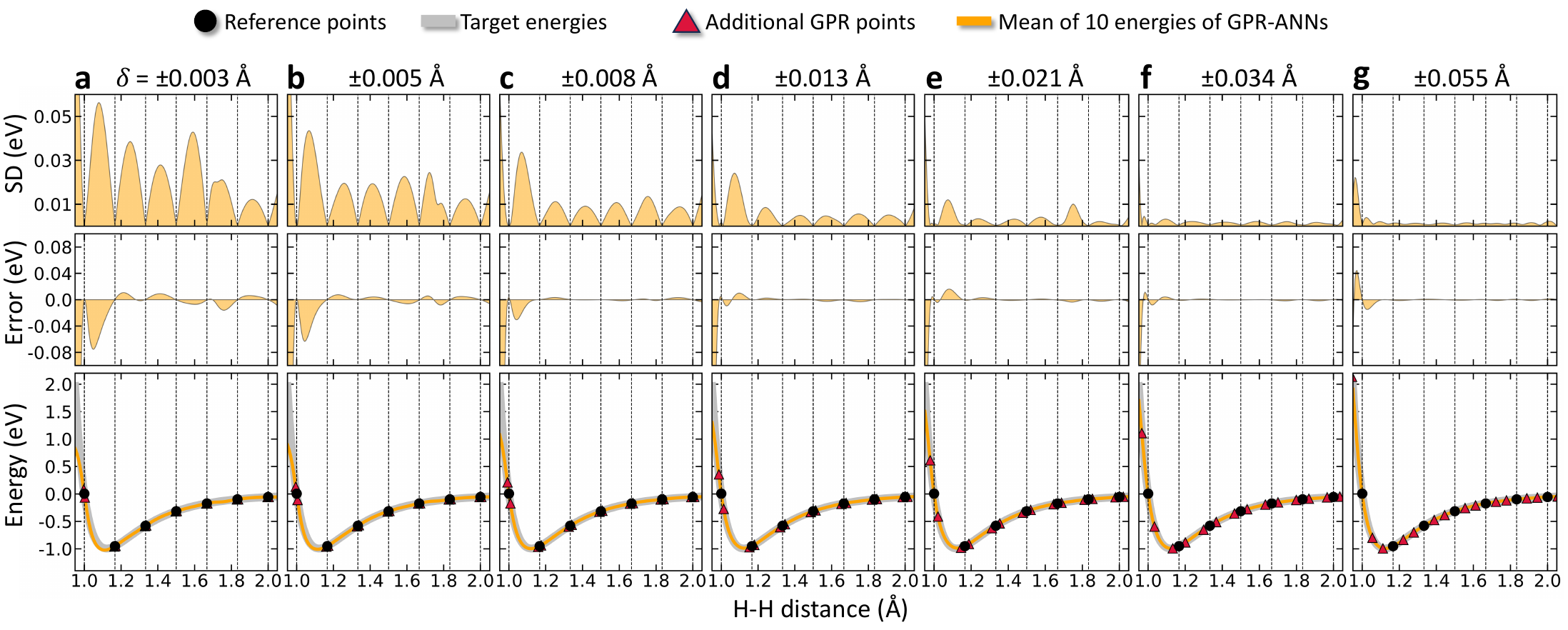}
  \caption{
  Standard deviation (SD) and error of energy predictions over a committee of 10 ANN potentials obtained from indirect force training with the GPR-ANN method. The potentials were trained on the seven reference points (black circles) and 14 GPR-augmented energies (red triangles) with different displacement amplitudes: \textbf{a} $\pm{}0.003$~Å, \textbf{b} $\pm{}0.005$~Å, \textbf{c} $\pm{}0.008$~Å, \textbf{d} $\pm{}0.013$~Å, \textbf{e} $\pm{}0.021$~Å, \textbf{f} $\pm{}0.034$~Å, and \textbf{g} $\pm{}0.055$~Å. The error is defined as the difference between the mean predicted energies (solid orange line) and the target potential energies (thick gray line) shown on the bottom panels.}
\end{figure}

\begin{figure}[H]
  \centering
  \includegraphics[width=1.0\textwidth]{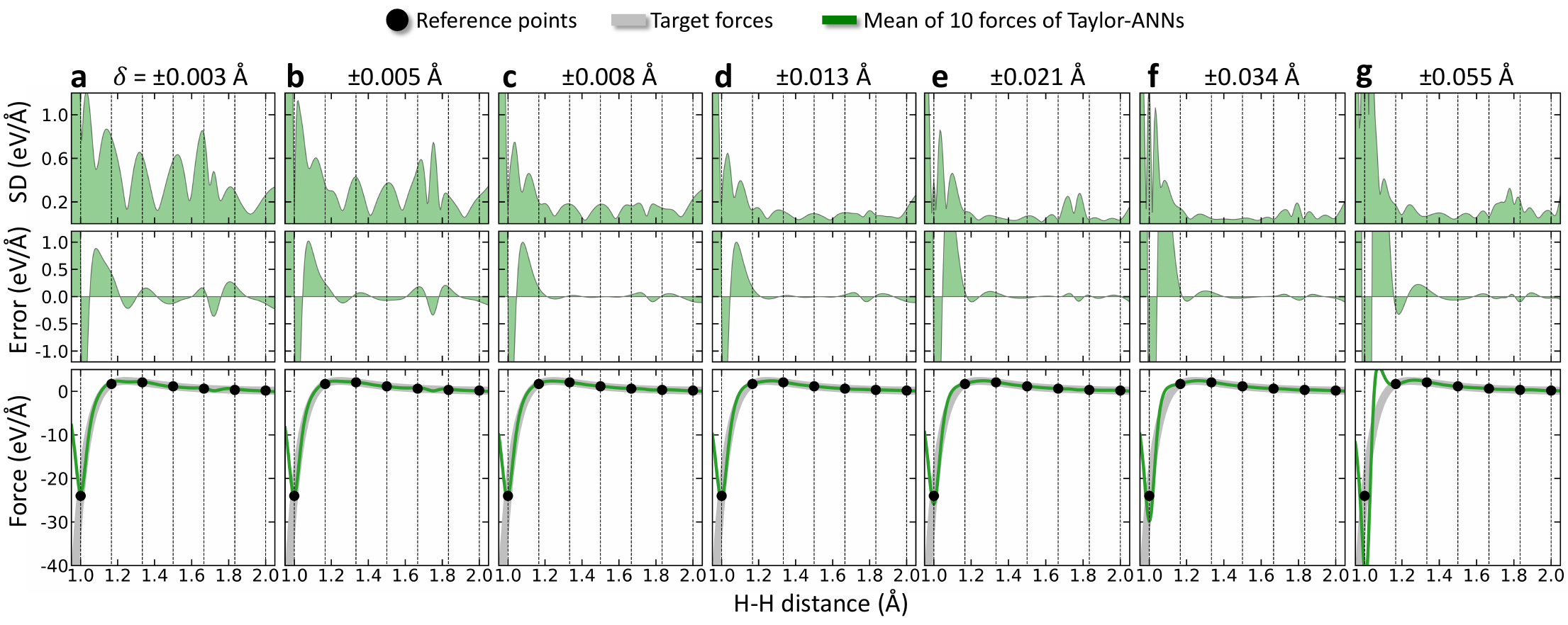}
  \caption{
  \textbf{a--g} Standard deviation (SD) and error of force predictions over a committee of 10 ANN potentials obtained from the Taylor-ANN training with different displacement amplitudes. The error is the difference between the mean predicted forces (solid green line) and the target potential forces (thick gray line) shown on the bottom panels.}
\end{figure}

\begin{figure}[H]
  \centering
  \includegraphics[width=1.0\textwidth]{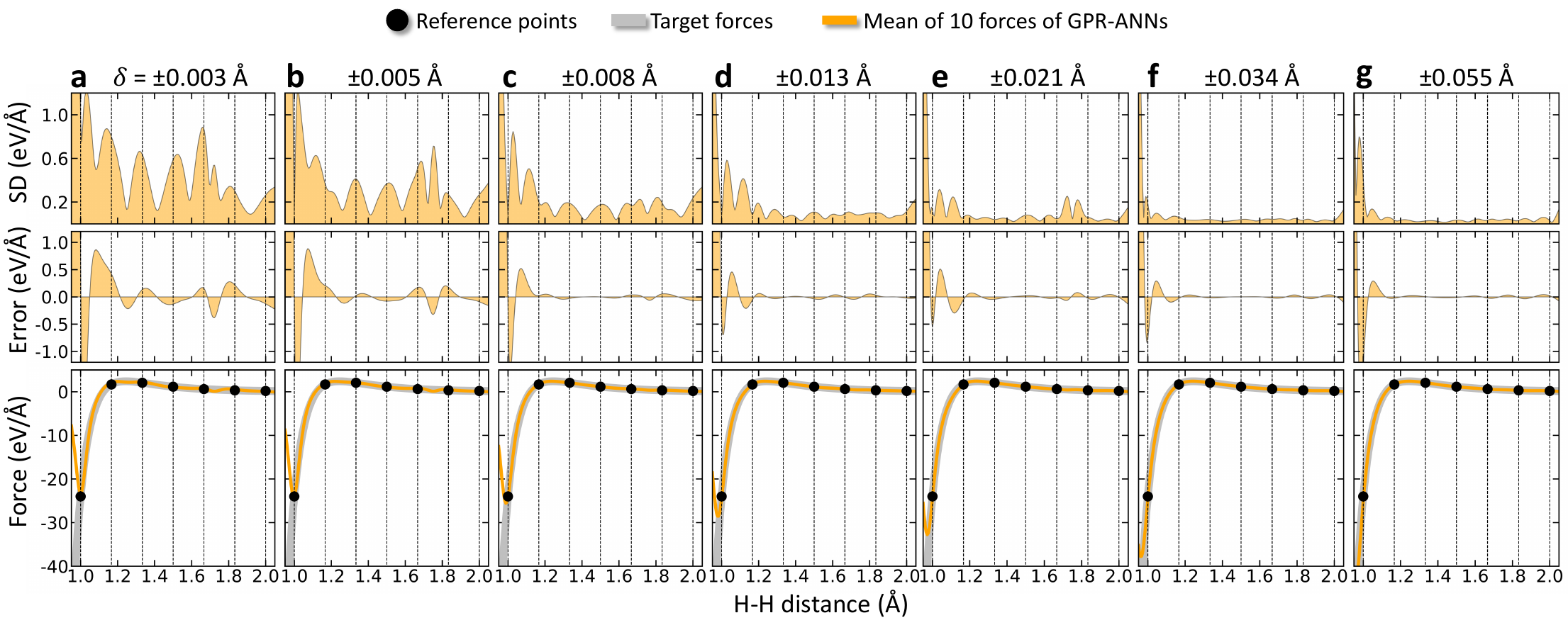}
  \caption{
  \textbf{a--g} Standard deviation (SD) and error of force predictions over a committee of 10 ANN potentials obtained from the GPR-ANN training with different displacement amplitudes. The error is the difference between the mean predicted forces (solid orange line) and the target potential forces (thick gray line) shown on the bottom panels.}
\end{figure}

\begin{figure}[H]
  \centering
  \includegraphics[width=0.7\textwidth]{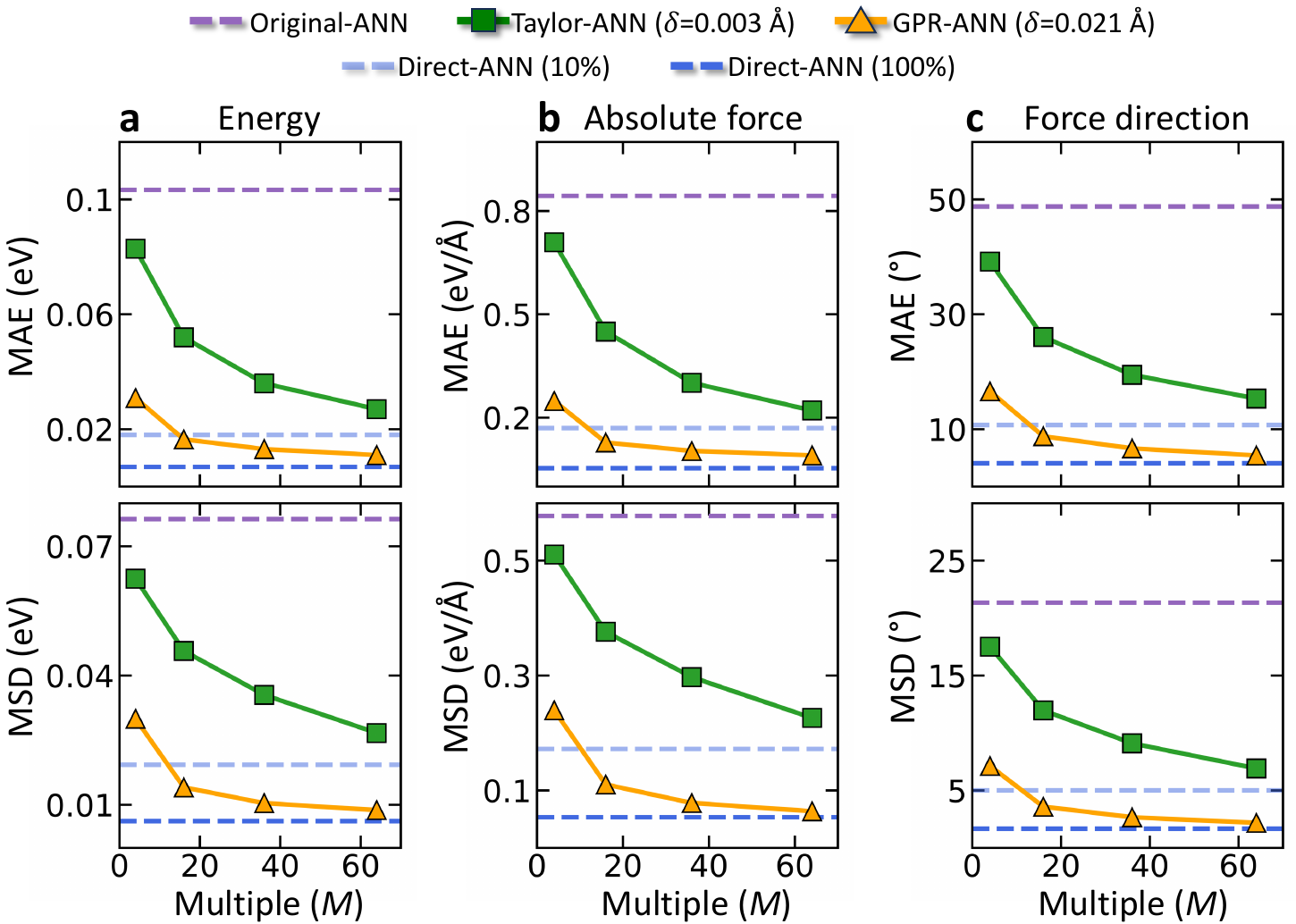}
  \caption{
   \textbf{The accuracy and robustness of the four ANN training methods for ethylene carbonate dimer structures.} The mean absolute error (MAE) and mean standard deviation (MSD) over a committee of 10 ANN potentials are shown for \textbf{a}~the energy, \textbf{b}~the absolute force magnitude, and \textbf{c}~the force direction as a function of multiple $M$, a parameter for the data augmentation methods. These metrics are shown for ANN potentials obtained from energy-only training (dashed purple line), indirect force training with the Taylor-ANN (green squares), and the GPR-ANN (orange triangles) approach, and direct force training with 10\% forces (dashed light blue line) and 100\% force information (dashed dark blue line).}
\end{figure}

\begin{figure}[H]
  \centering
  \includegraphics[width=0.7\textwidth]{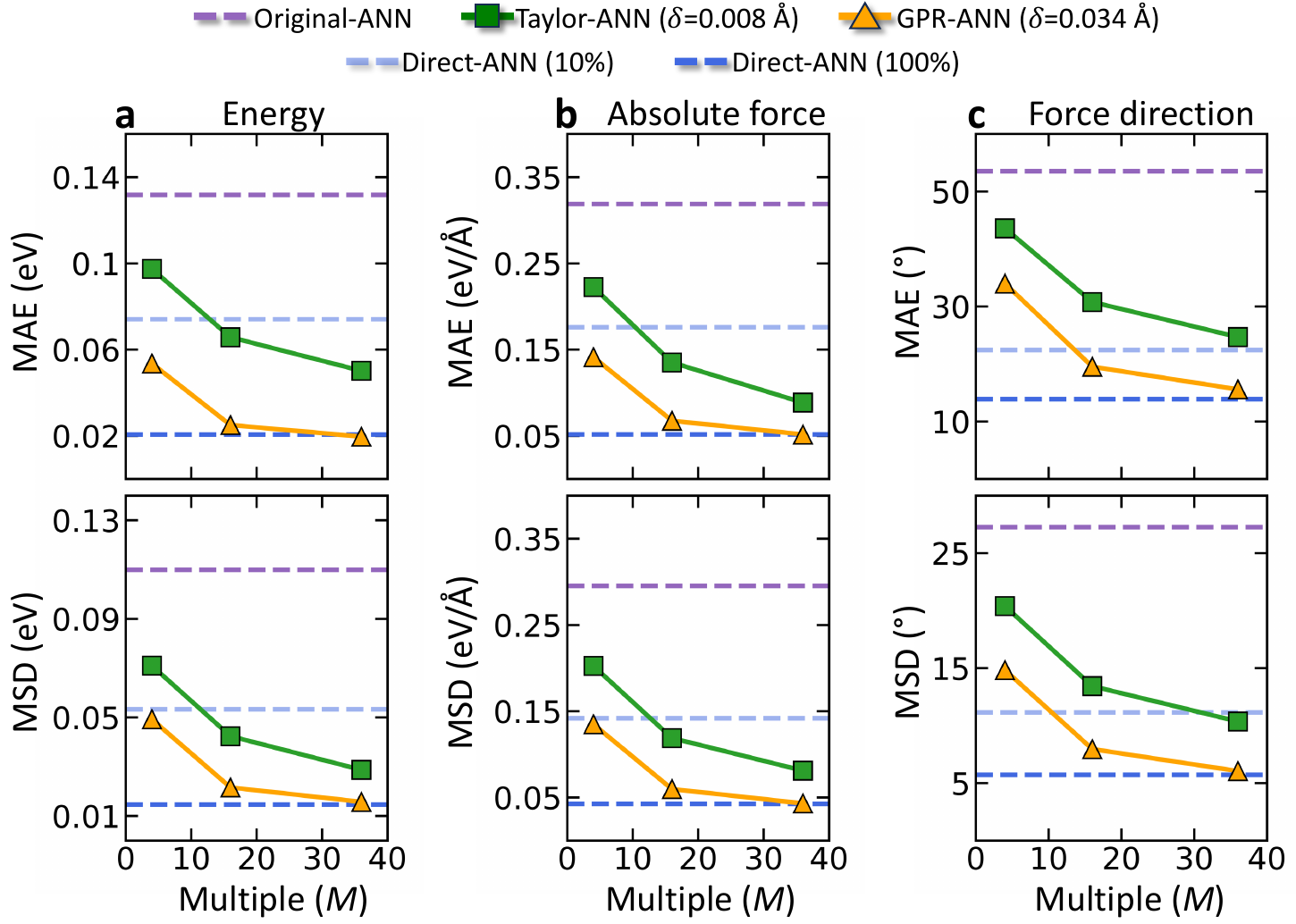}
  \caption{
   \textbf{The accuracy and robustness of the four ANN training methods for the database of an ethylene carbonate molecule adsorbed on the lithium metal (100) surface.} The mean absolute error (MAE) and mean standard deviation (MSD) over a committee of 10 ANN potentials are shown for \textbf{a}~the energy, \textbf{b}~the absolute force magnitude, and \textbf{c}~the force direction as a function of multiple $M$, a parameter for the data augmentation methods. These metrics are shown for ANN potentials obtained from energy-only training (dashed purple line), indirect force training with the Taylor-ANN (green squares), and the GPR-ANN (orange triangles) approach, and direct force training with 10\% forces (dashed light blue line) and 100\% force information (dashed dark blue line).}
\end{figure}

\begin{figure}[H]
  \centering
  \includegraphics[width=1.0\textwidth]{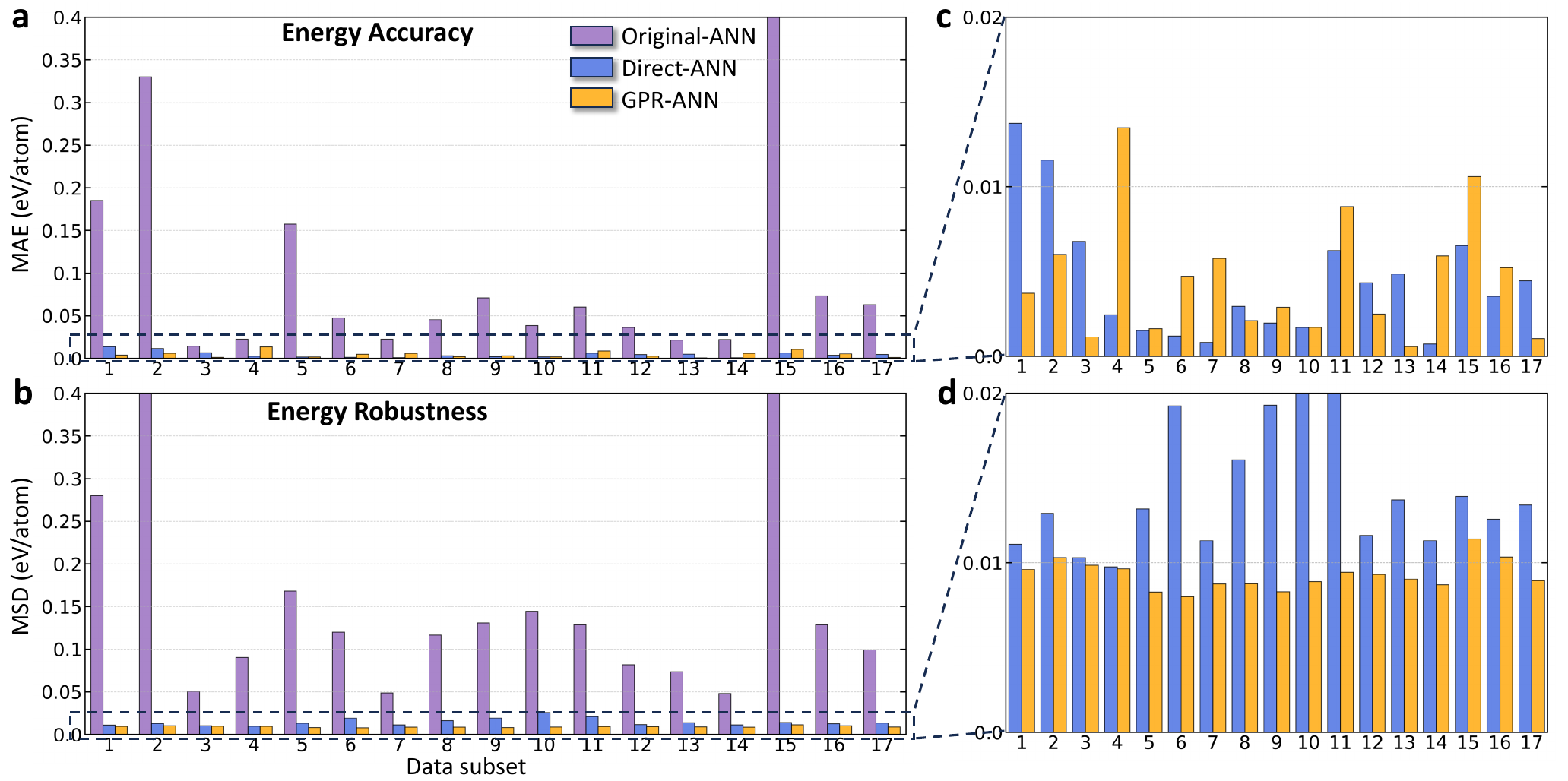}
  \caption{
   \textbf{Comparison of GPR-ANN and direct force training strategies for heterogeneous data of Li-EC interfaces.} \textbf{a} The mean absolute error (MAE) and \textbf{b} mean standard deviation (MSD) over a committee of 10 ANN potentials from direct force training (blue bar) and GPR-ANN training (orange bar) are shown for energy predictions on the test data of each data subset. For comparison, the MAE and MSD of ANN potentials from energy-only training are shown by purple bar. Zoomed-in \textbf{c} MAE and \textbf{d} MSD of the region marked with dashed rectangles. All the metrics are characterized with the optimal parameters for force trainings, i.e., $\delta=0.034$~Å and $M=36$ for GPR-ANN training and 100\% forces and $\alpha=0.3$ for direct force training.}
\end{figure}

\begin{figure}[H]
  \centering
  \includegraphics[width=1.0\textwidth]{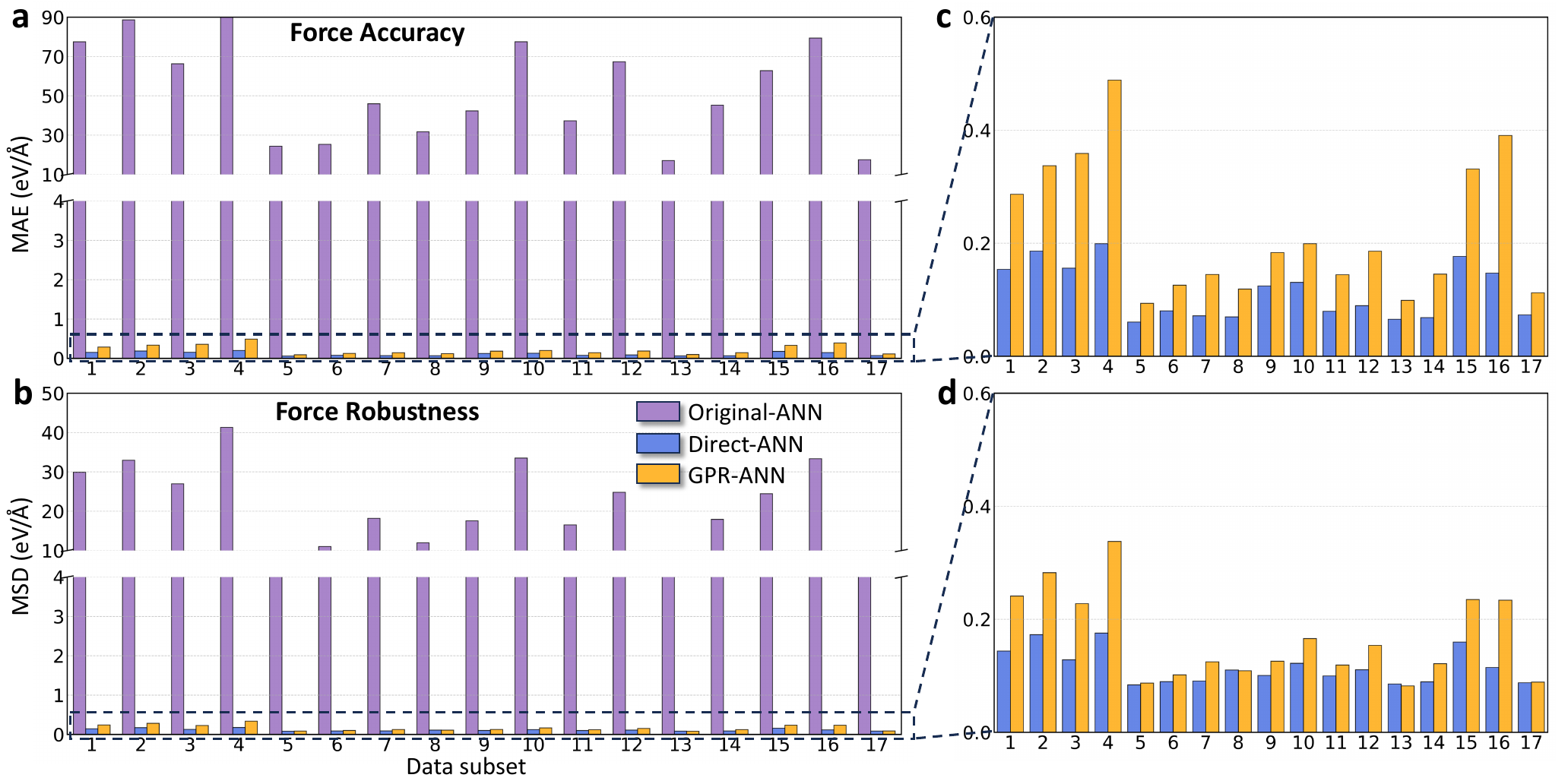}
  \caption{
   \textbf{Comparison of GPR-ANN and direct force training strategies for heterogeneous data of Li-EC interfaces.} \textbf{a} The mean absolute error (MAE) and \textbf{b} mean standard deviation (MSD) over a committee of 10 ANN potentials from direct force training (blue bar) and GPR-ANN training (orange bar) are shown for force predictions on the test data of each data subset. For comparison, the MAE and MSD of ANN potentials from energy-only training are shown by purple bar. Zoomed-in \textbf{c} MAE and \textbf{d} MSD of the region marked with dashed rectangles. All the metrics are characterized with the optimal parameters for force trainings.}
\end{figure}

\begin{figure}[H]
  \centering
  \includegraphics[width=0.4\textwidth]{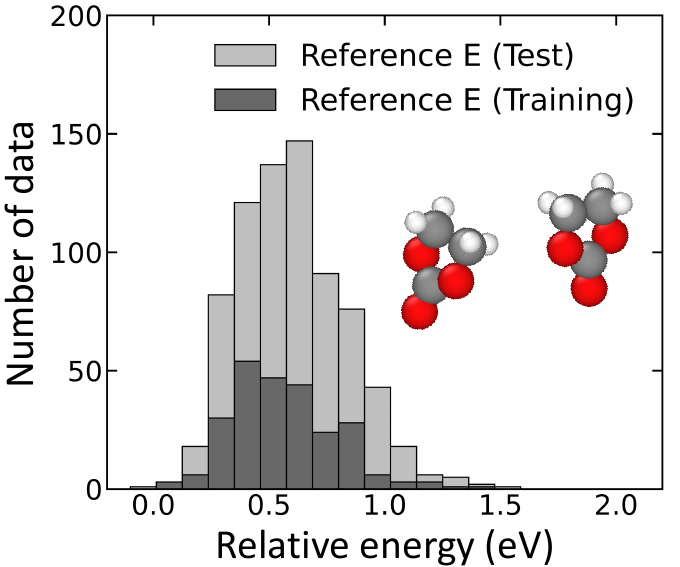}
  \caption{
   Energy distribution of ethylene carbonate dimer database with respect to the minimum energy. The energy and force of the reference 1,000 structures were calculated using HSE06 DFT calculations, and they were divided into 250 training (dark gray bar) and 750 test data (light gray bar).}
\end{figure}

\begin{figure}[H]
  \centering
  \includegraphics[width=0.8\textwidth]{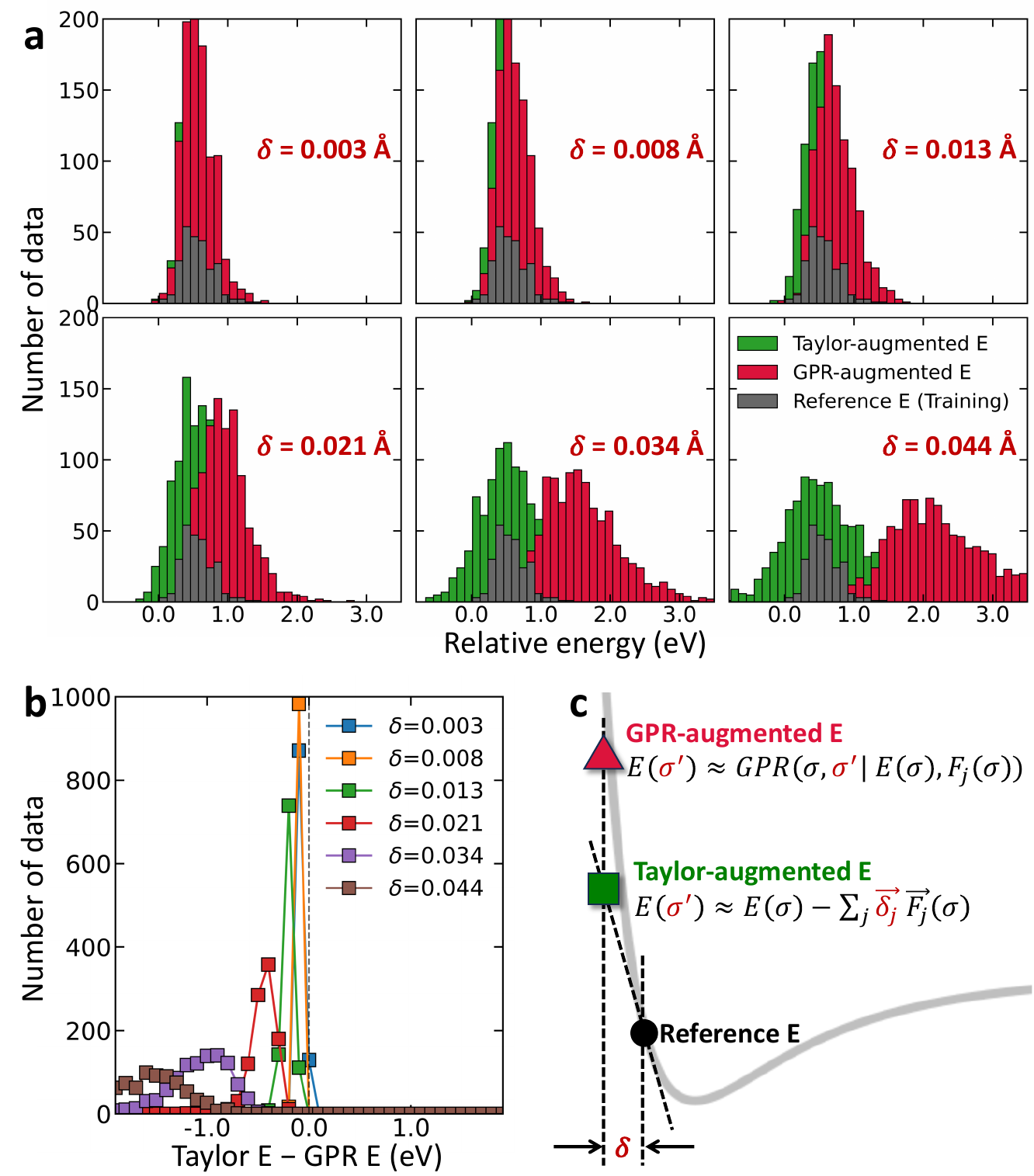}
  \caption{
   \textbf{a} Distribution of synthetic energy data from the linear Taylor expansion (green bar) and GPR model (red bar) for the additional structures generated by randomly displacing the 250 reference training data (dark gray bar) for ethylene carbonate dimer. \textbf{b} Difference between Taylor- and GPR-augmented energies for different displacement length $\delta$. \textbf{c} A schematic of Taylor- (green square) and GPR-approximate energy (red triangle) around potential energy surface with positive curvature.}
\end{figure}

\begin{figure}[H]
  \centering
  \includegraphics[width=0.4\textwidth]{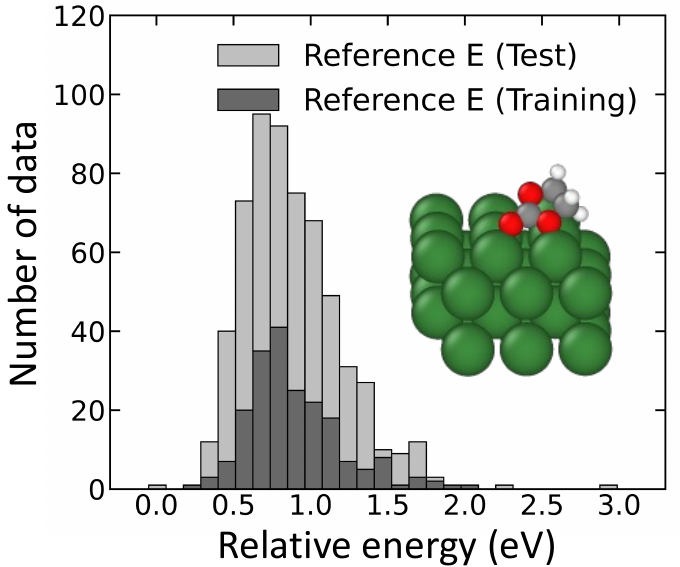}
  \caption{
   Energy distribution of the reference data for an ethylene carbonate molecule adsorbed on the lithium metal (100) surface. The energy and force of the reference 800 structures were calculated using HSE06 DFT calculations with 5$\times$5$\times$1 k-point meshes, and they were divided into 200 training (dark gray bar) and 600 test data (light gray bar).}
\end{figure}

\begin{figure}[H]
  \centering
  \includegraphics[width=0.8\textwidth]{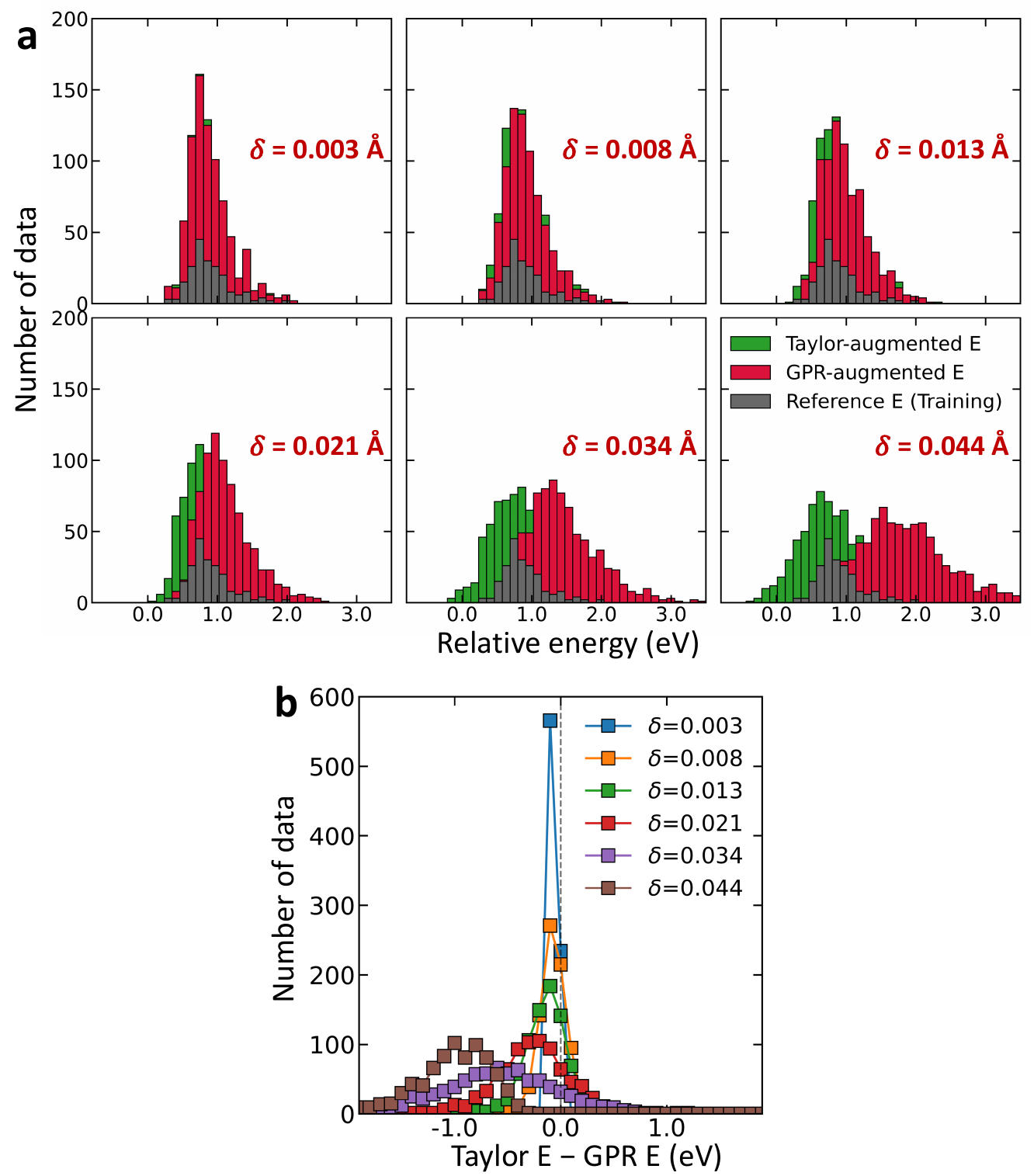}
  \caption{
   \textbf{a} Distribution of synthetic energy data from the linear Taylor expansion (green bar) and GPR model (red bar) for the additional structures generated by randomly displacing the 200 reference training data (dark gray bar) for an ethylene carbonate molecule adsorbed on the lithium metal (100) surface. \textbf{b} Difference between Taylor- and GPR-augmented energies for different displacement length $\delta$.}
\end{figure}

\begin{figure}[H]
  \centering
  \includegraphics[width=0.8\textwidth]{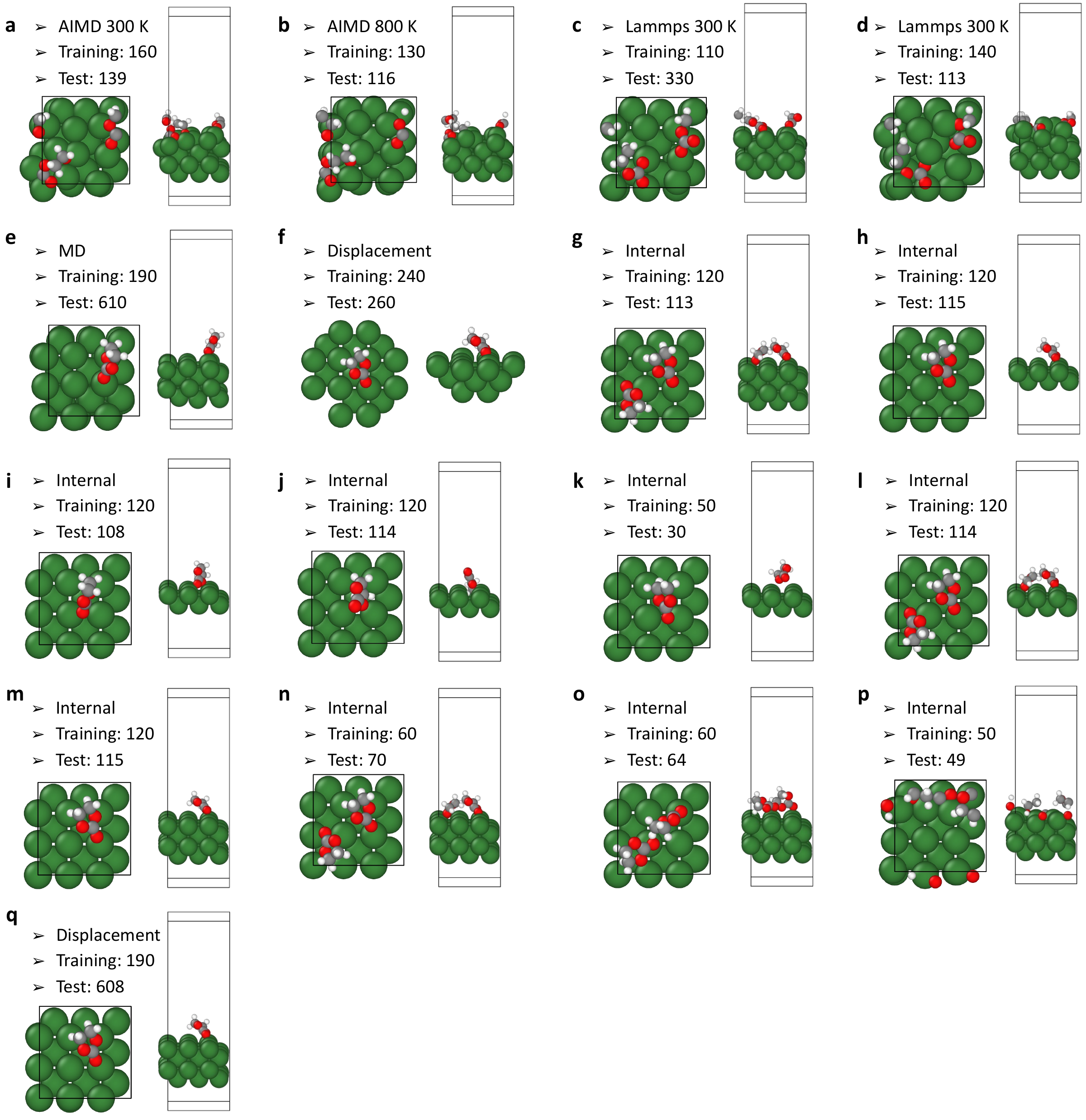}
  \caption{
   Top and side view of a representative atomic structure, sampling method, and the number of training and test reference structures for each subset of heterogeneous database for Li-EC interfaces.}
\end{figure}

\begin{figure}[H]
  \centering
  \includegraphics[width=0.6\textwidth]{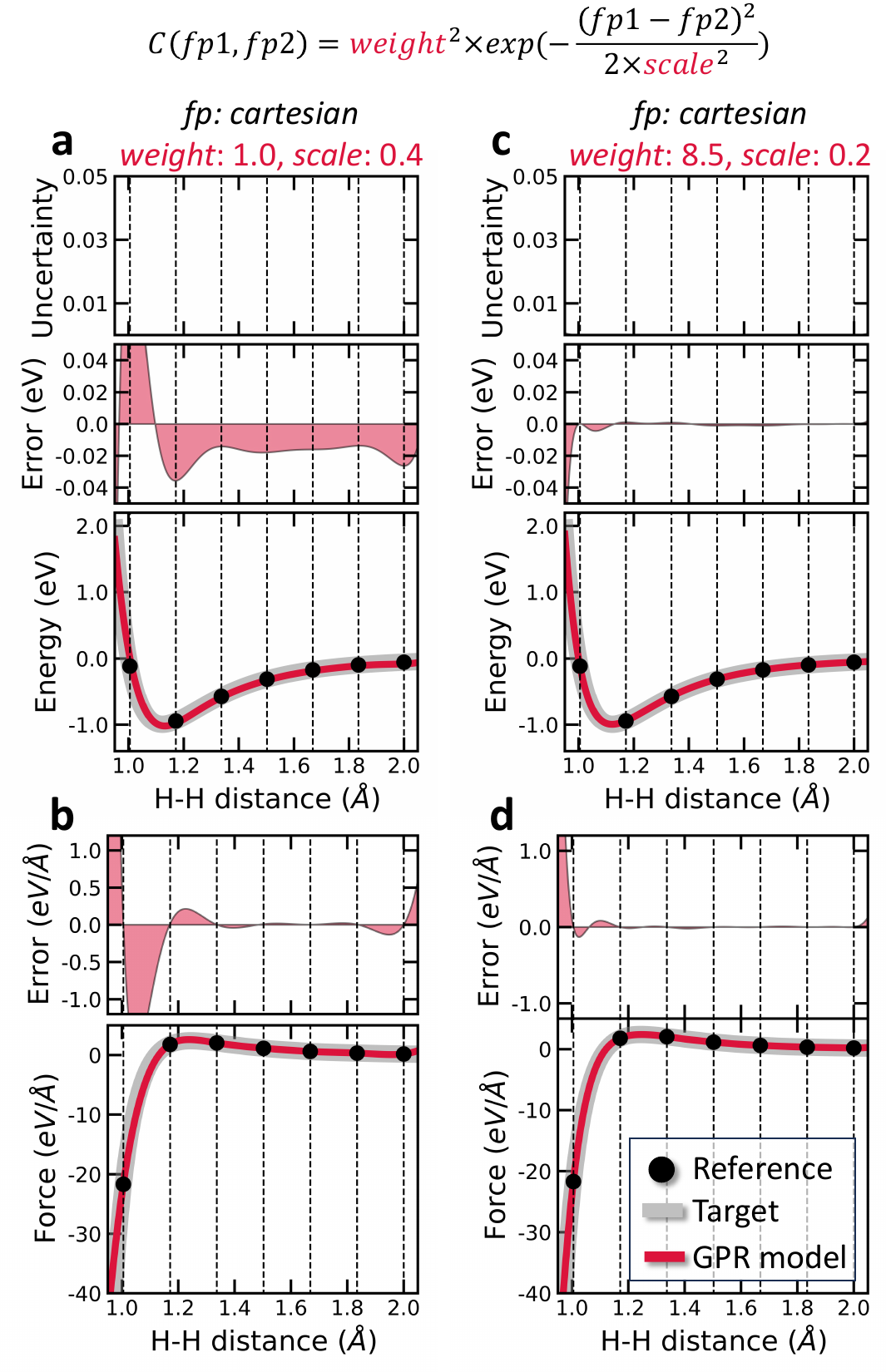}
  \caption{
   \textbf{GPR model for an \ce{H2} molecule.} The energy and force of seven reference samples (black circles) were used to construct a GPR model (red line) with cartesian fingerprint. Uncertainty and error of GPR \textbf{a} energy and \textbf{b} force predictions with default kernel parameters. After the kernel parameter optimization, \textbf{c} energy and \textbf{d} force predictions by the GPR model align more closely with the target values (thick gray line).}
\end{figure}

\begin{figure}[H]
  \centering
  \includegraphics[width=0.6\textwidth]{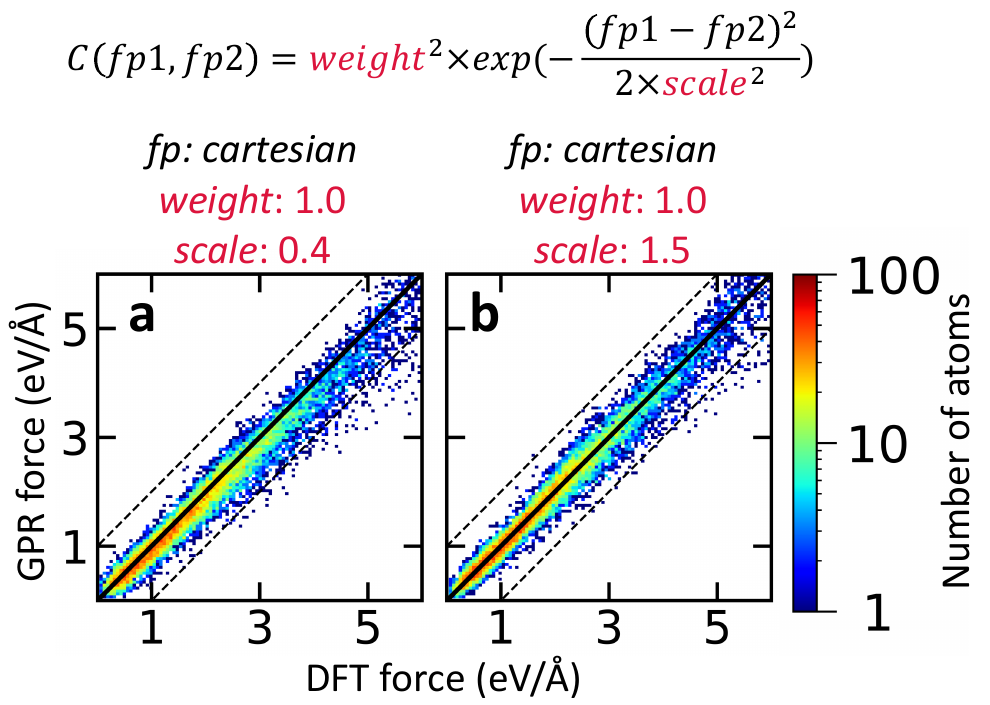}
  \caption{
   \textbf{GPR model for ethylene carbonate dimers.} The energy and force of 250 training samples were used to construct a GPR model with cartesian fingerprint. Correlation between the DFT reference and the predicted absolute force by GPR model with kernel parameters \textbf{a} before and \textbf{b} after the optimization.}
\end{figure}

\begin{figure}[H]
  \centering
  \includegraphics[width=1.0\textwidth]{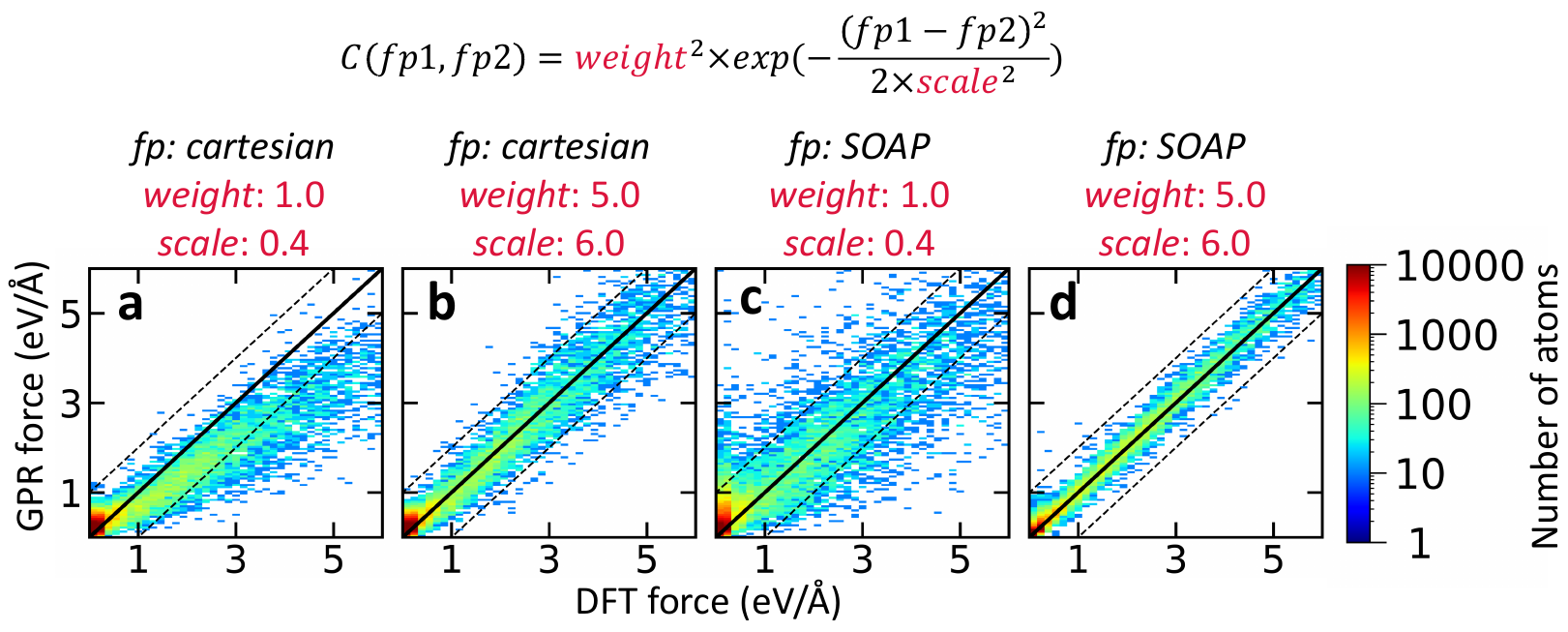}
  \caption{
   \textbf{GPR model for the database of an ethylene carbonate molecule adsorbed on the lithium metal (100) surface.} The energy and force of 200 training samples were used to construct a GPR model with cartesian and SOAP fingerprints. Correlation between the DFT reference and the predicted absolute force by GPR model with the cartesian fingerprint \textbf{a} before and \textbf{b} after the kernel parameter optimization. The absolute force correlation with the SOAP fingerprint \textbf{c} before and \textbf{d} after the kernel parameter optimization.}
\end{figure}

\end{document}